\documentclass[draftcls,journal,onecolumn]{IEEEtran}

\usepackage{cite}
\usepackage{times}
\usepackage[named]{algo}
\usepackage{amsmath}
\usepackage{amssymb}
\usepackage[pdftex]{graphicx}
\usepackage[small]{caption}
\usepackage{subfigure}
\usepackage{url}

\graphicspath{{./}{images/}}

\newtheorem{theorem}{Theorem}
\newtheorem{lemma}[theorem]{Lemma}

\newtheorem{corollary}[theorem]{Corollary}
\newtheorem{Definition}{Definition}

\newtheorem{proposition}[theorem]{Proposition}
\newtheorem{problem}{Problem}

\newcommand{\eat}[1]{}

\newcommand{\hide}[1]{}
\newcommand{\junk}[1]{}
\newcommand{\etal}{ {\em et. al. } }

\newcommand{\opt}{\mbox{O{\sc pt}}}

\newcommand{\vect}{\boldsymbol}

\newcommand{\psia}{\psi^{a}}
\newcommand{\psib}{\psi^{b}}

\newcommand{\phia}{\phi^{a}}

\newcommand{\E}{{\cal E}}

\newcommand{\wai}{{\cal W}^{-1}}

\newcommand{\real}{{\cal R}}

\providecommand{\abs}[1]{\lvert#1\rvert}
\providecommand{\norm}[1]{\lVert#1\rVert}
\providecommand{\snorm}[1]{\bigl\lVert#1\bigr\rVert}

\newcommand{\fp}{f^*_{(p)}}
\newcommand{\fin}{f^*_{(\infty)}}
\newcommand{\hfi}{\hat{f}_{(\infty)}}
\newcommand{\hfo}{\hat{f}_{(1)}}
\newcommand{\fo}{f^*_{(1)}}
\newcommand{\q}{\ensuremath{{p'}}}
\newcommand{\taps}{\ensuremath{q}}

\begin{document}
\date{}
\title{Approximation Algorithms for Wavelet Transform Coding of Data
  Streams\thanks{A preliminary and significantly weaker version of this paper appeared as an
    extended abstract in the conference SODA 2006~\cite{GH06}.}}
\author{Sudipto~Guha and Boulos~Harb,~\IEEEmembership{Student Member,~IEEE}
\thanks{This work was supported in part by an
Alfred P. Sloan Research Fellowship and by an NSF Awards CCF-0430376 and CCF-0644119.}
\thanks{S. Guha and B. Harb are with the 
Department of Computer Information Science,
University of Pennsylvania,
3330 Walnut St,Philadelphia, PA 19104.
(email: \{sudipto,boulos\}@cis.upenn.edu).}
}

\maketitle

\begin{abstract} 
%\boldmath{
This paper addresses the problem of finding a $B$-term wavelet
representation of a given discrete function $f \in \real^n$ whose
distance from $f$ is minimized. The problem is well understood when we
seek to minimize the Euclidean distance between $f$ and its
representation. The first known algorithms for finding provably
approximate representations minimizing general $\ell_p$ distances
(including $\ell_\infty$) under a wide variety of compactly supported
wavelet bases are presented in this paper. For the Haar basis, a
polynomial time approximation scheme is demonstrated. These algorithms
are applicable in the one-pass sublinear-space data stream model of
computation. They generalize naturally to multiple dimensions and
weighted norms.  A universal representation that provides a provable 
approximation guarantee under all $p$-norms simultaneously;  
and the first approximation algorithms for bit-budget
versions of the problem, known as adaptive quantization, are also presented. 
Further, it is shown that the algorithms presented here can be used to select 
a basis from a tree-structured dictionary of bases and find a $B$-term
representation of the given function that provably approximates its
best dictionary-basis representation.
%}
\end{abstract}

\begin{keywords}
Nonlinear approximation, compactly supported
wavelets, transform coding, best basis selection.
\end{keywords}

A central problem in approximation theory is to represent a function
concisely. Given a function or a signal as input, the goal is to
construct a representation as a linear combination of several
predefined functions, under a constraint which limits the space used
by the representation. The set of predefined functions are denoted as
the dictionary. One of the most celebrated approaches in this context
has been that of \emph{nonlinear approximation}.  In this approach the
dictionary elements that are used to represent a function are allowed
to depend on the input signal itself.

Nonlinear approximations has a rich history starting from the work of
Schmidt \cite{schmidt}; however, more recently these have come to fore
in the context of wavelet dictionaries \cite{wave2,wave3}.  Wavelets
were first analyzed by DeVore~\etal~\cite{devoreWavelets} in nonlinear
approximation. Wavelets and multi-fractals have since found extensive
use in image representation, see Jacobs \cite{wavei}. In fact, the
success of wavelets in nonlinear approximation has been hailed by many
researchers as ``the `true' reason of the usefulness of wavelets in
signal compression'' (Cohen~\etal~\cite{cohen97importance}). Due to
lack of space we would not be able to review the extremely rich body
of work that has emerged in this context; see the surveys by DeVore
\cite{devore} and Temlyakov~\cite{T03} for substantial reviews.

\bigskip However, with the rise in the number of domains for which
wavelets have been found useful, several interesting problems have
arisen. Classically, the error in terms of representation has been
measured by the Euclidean or $\ell_2$ error.  This choice is natural for
analysis of functions, but not necessarily for representation of data and
distributions. Even in image compression, Mallat~\cite[p.~528]{wave2} 
and Daubechies~\cite[p.~286]{wave3} point out that while the $\ell_2$ 
measure does not adequately quantify perceptual errors, it is used, 
nonetheless, since other norms are difficult to optimize. 
However, non-$\ell_2$ measures have been widely used in the
literature.  Matias, Vitter and Wang \cite{MVW98}, suggested using the 
$\ell_1$ metric and showed that wavelets could be used in creating 
succinct synopses of data allowing us to answer queries approximately. 
The $\ell_1$ distance is a statistical distance and is well suited for
measuring distributions.  Interestingly, Chapelle, Haffner and 
Vapnik~\cite{CHV} show that the $\ell_1$ norm significantly outperforms the
$\ell_2$ norm in image recognition on images in the Corel data set using
SVM's. From a completely different standpoint, we may be interested in
approximating a signal in the $\ell_\infty$ norm thus seeking a high
fidelity approximation throughput rather than an `average' measure such as
other norms. This is particularly of interest if we are trying to process
noisy data (we consider $\ell_1,\ell_\infty$ approximations in Section~
\ref{sec:imagecompress}).
While we have developed a reasonable understanding of $\ell_2$
error, problems involving non-$\ell_2$ error are still poorly understood.
This paper takes the first steps towards filling this gap.

\bigskip One of the most basic problems in nonlinear approximation is
the following: Given a wavelet basis $\{\psi_i\}$ and a target
function (or signal, vector) $f\in\real^n$, construct a representation
$\hat{f}$ as a linear combination of at most $B$ basis vectors so as
to minimize some normed distance between $f$ and $\hat{f}$.  The
\emph{$B$-term representation} $\hat{f}$ belongs to the space
$\mathcal{F}_B = \{\sum_{i=1}^n z_i \psi_i : z_i\in\real, \norm{z}_0
\le B\}$, where $\norm{z}_0$ is the number of non-zero coefficients in
$z\in\real^n$.  The problem is well-understood if the error of the
representation is measured using the Euclidean or $\ell_2$ distance.
Since the $\ell_2$ distance is preserved under rotations, by
Parseval's theorem, we have
\[ 
\norm{f - \hat{f}}^2_2 = 
\sum_i \left( f_i -\sum\nolimits_j z_j\psi_j[i] \right)^2=
\sum_i \left( \langle f,\psi_i\rangle - z_i \right)^2 \enspace .
\]
It is clear then that the solution under this error measure is to
retain the largest $B$ inner products $ \langle f,\psi_i\rangle$,
which are also the coefficients of the wavelet expansion of $f$.  {\em Note:}
the fact that we have to store the inner products or the wavelet coefficients
is a natural consequence of the proof of optimality.

The common strategy for the $B$-term representation problem in the
literature has been ``to retain the [$B$] terms in the wavelet
expansion of the target function which are largest relative to the
norm in which error of approximation is to be measured''
\cite[p. 4]{devore}.  This strategy is reasonable in an extremal
setting; i.e., if we are measuring the rate of the error as a function
of $B$.  
\hide{An example of such a result is~\cite[p.~397]{wave2}: If
$f$ is bounded variation, then the nonlinear approximation of $f$
using its largest wavelet coefficients decays like
$(\norm{f}_{V}/B)^2$. Here $\norm{f}_V = \int_0^1 \abs{f'(t)}
\mathit{dt}$ is the total variation norm of $f$.} %
But it is easy to show that the common greedy strategy is sub-optimal,
see \cite{GG-TODS,GK05,muthu-wave,MU,yossi1,G05}. In light of this,
several researchers \cite{GK04,GK05,muthu-wave,MU,G05} considered a
{\em restricted} version of the problem under the Haar basis where we
may only choose wavelet coefficients of the data. However to date, the
only bound on its performance with respect to the target function's
best possible representation using $B$ terms from the wavelet basis is
given by Temlyakov~\cite{T98} (see also~\cite[Sec.~7]{T03}).
Temlyakov shows that given $f$ in the (infinite dimensional) Banach
function space $L_p[0,1]$, $1 < p < \infty$, if the given basis
$\{\psi_i\}_{i\in\mathbb{Z}}$ is \emph{$L_p$-equivalent} to the Haar
basis~\cite{DKT98}, then the error of the common greedy strategy is an
$\alpha$ factor away from that of the optimal $B$-term
representation. The factor $\alpha$ depends on $p$ and properties of
$\{\psi_i\}$, but the dependence is unspecified.  
\hide{It is noteworthy that compactly supported wavelet bases are
$L_p$-equivalent to the Haar basis~\cite{DKT98}.} % 
However, from an optimization point of view in the finite-dimensional
setting, the relationship between the factor $\alpha$ and the
dimension $n$ of the space spanned is the key problem, which we
address here.  Three relevant questions arise in this context.  First
is whether there are universal algorithms/representations that
simultaneously approximate all $\ell_p$ norms. This is important
because in many applications, it is difficult to determine the most
suitable norm to minimize without looking at the data, and an
universal representation would be extremely useful. The second
question concerns the complexity of representing the
optimal solution.  It is not immediate {\it a priori} that the optimal
unrestricted solution minimizing, for example, the $\ell_5$ norm for a
function that takes only rational values can be specified by $B$
rational numbers. The third related question pertains to the
computational complexity of finding the optimum solution. Can the
solution be found in time polynomial in the size of the input $n$? Or
better yet, can the solution be found in {\em strongly} polynomial
time where the running time of the algorithm does not depend on the
numeric values of the input.  We focus on these questions using the
lens of {\em approximation algorithms}, where we seek to find a
solution that is close to the optimum---in fast polynomial time.  Note
that the use of approximation algorithms does not limit us from using
additional heuristics from which we may benefit, but gives us a more
organized starting point to develop heuristics with provable bounds.

\bigskip
A natural generalization of the problem above is known as 
{\em Adaptive Quantization}. The $B$-term representation requires 
storing $2B$ numbers, the coefficient and the index of the corresponding basis vector to be
retained.  The actual cost (in bits) of storing the real numbers $z_i$
is, however, non-uniform. Depending on the scenario, it may be
beneficial to represent a function with a large number of low-support
vectors with low precision $z_i$'s or a few vectors with more detailed
precision $z_i$'s.  Hence, a $B$-term representation algorithm does
not translate directly into a practical compression algorithm.  A
natural generalization, and a more practical model as noted in
\cite{cohen97importance}, is to minimize the error subject to the
constraint that the stored values and indices cannot exceed a given
bit-budget.  Note that, again, we are not constrained here to storing
wavelet expansion coefficients. This bit-budget version of the problem
is known as adaptive quantization, which we will also consider.
To the best of our knowledge, there are no known approximation
algorithms for this problem.

\bigskip
One other natural generalization incorporates a choice of basis into
the optimization problem~\cite{devore}.  We are given a dictionary
$\mathcal{D}$ of bases and our objective is to choose a best basis in
$\mathcal{D}$ for representing $f$ using $B$ terms.  This bi-criteria
optimization problem is a form of \emph{highly nonlinear
approximation}~\cite{devore}.  In a seminal work, Coiffman and
Wickerhauser~\cite{CW92} construct a binary tree-structured dictionary
composed of $O(n\log n)$ vectors and containing $2^{O(\frac{n}{2})}$
orthonormal bases. They present a dynamic programming algorithm that
in $O(n\log n)$ time finds a best basis minimizing the entropy of its
inner products with the given function $f$.  Mallat~\cite{wave2}
discusses generalizations based on their algorithm for finding a basis
from the tree dictionary that minimizes an arbitrary concave function
of its expansion coefficients.  However, finding a basis in
$\mathcal{D}$ that minimizes a concave function of its inner products
with the given $f$ is not necessarily one with which we can best
represent $f$ (in an $\ell_p$ sense) using $B$ terms.  Combining our
approximation algorithms for the original $B$-term representation
problem with the algorithm of Coiffman and Wickerhauser, we show how
one can construct provably-approximate $B$-term representations in
tree-structured wavelet dictionaries. Several of these results also extend
to arbitrary dictionaries with low coherence~\cite{MallatNPhrad,redundant}.
\eat{
Another form of highly nonlinear approximation, and a further
generalization, allows for representations using vectors from the
dictionary that do not necessarily form a basis.  In an arbitrary setting,
this problem is NP-hard via a reduction from exact Set Cover as shown by
Davis, Mallat and Avellaneda \cite{MallatNPhrad}.  In fact, and as noted by
Gilbert, Muthukrishnan, and Strauss \cite{redundant}, the same proof shows
that even approximating the error of the optimal $B$-term representation
up to any constant factor is NP-hard. Thus the natural avenue is to
investigate more structured dictionaries that arise in practice, and in
particular the tree dictionaries of Coiffman and Wickerhauser \cite{CW92}.
Our results in this paper therefore separate the tree structured
dictionaries and these arbitrary dictionaries from the perspective of
approximation.
We note that Gilbert, Muthukrishnan, and Strauss \cite{redundant} focus on
redundant dictionaries with small coherence (that is, the vectors in the
dictionary are nearly orthogonal). They construct $(1+\epsilon)$
approximate representations under the $\ell_2$ error measure in any such
incoherent dictionary. They list the problem of representation using
redundant dictionaries under general $\ell_p$ norms as an open question.
Further, their results do not apply to the tree-structured dictionaries we
consider, since these dictionaries are highly coherent.
}%

\bigskip Along with the development of richer representation
structures, in recent years there has been significant increase in the
data sets we are faced with. At these massive scales, the data is not
expected to fit the available memory of even fairly powerful
computers. One of the emergent paradigms to cope with this challenge
is the idea of \emph{data stream algorithms}. In a data stream model
the input is provided one at a time, and any input item not explicitly
stored is inaccessible to the computation, i.e., it is lost. The
challenge is to perform the relevant computation in space that is
sublinear in the input size; for example, computing the best
representation of a discrete signal $f(i)$ for $i \in [n]$ that is
presented in increasing order of $i$, in only $o(n)$ space. This is a
classic model of time-series data, where the function is presented one
value at a time.  It is immediate that under this space restriction we
may not be able to optimize our function. This harks back to the issue
raised earlier about the precision of the solution. Thus, the question
of approximation algorithms is doubly interesting in this context. 
The only known results on this topic \cite{GKMS01,GGIKMS02} crucially depend on
Parseval's Identity and do not extend to norms other than $\ell_2$.
\eat{
For the case of $\ell_2$ error, Gilbert~\etal~\cite{GKMS01} show that the
optimum representation can be achieved in a data stream setting. In a
significantly stronger model of data streams, where $f(i)$ is
specified by a series of updates, Gilbert~\etal~\cite{GGIKMS02} show
how to achieve $(1+\epsilon)$ approximation for $\ell_2$
error. Although both these papers state their results in terms of Haar
wavelets, the claims extend to arbitrary compactly supported wavelets
because the cores of both papers provide a way for maintaining (in the
more general model, only approximately) the largest $B$
terms. However, the critical part of both proofs is Parseval's
identity which does not hold under non-$\ell_2$ error
measures. \eat{Unfortunately, it is precisely those cases where the data is
large (and thus we benefit from a data stream computation), that we
need a non-$\ell_2$ error measure; for example, the $\ell_1$ measure
or variational distances for distributions.}
}%

\bigskip
In summary, even for the simplest possible transform coding problem,
namely the $B$-term representation problem, we can identify the
following issues:

\begin{itemize}
\item There are no analysis techniques for $\ell_p$ norms.  In fact
  this is the bottleneck in analyzing any generalization of the
  $B$-term representation problem; e.g., the adaptive quantization problem.
\item All of the (limited) analyses in the optimization setting have
  been done on the Haar system, which although important, is not the
  wavelet of choice in some applications. 
  Further, in this setting, the bounds on the performance of the
  algorithms used in practice which retain wavelet coefficients are unclear.
\item Signals that require transform coding are often presented as a
  streaming input---no algorithms are known except for $\ell_2$ norms.
\item The computational complexity of transform coding problems
  for structured dictionaries, or even for wavelet bases, is unresolved.
\end{itemize}

\subsection{Our Results} 
We ameliorate the above by showing:

\begin{enumerate}
\item For the {\em $B$-term representation problem} we show that, 
\begin{enumerate}
\item The restricted solution that retains at most $B$ \emph{wavelet
  coefficients} is a $O(\log n)$ approximation to the
  unrestricted solution under all $\ell_p$ distances for general compact
  systems (e.g. Haar, Daubechies, Symmlets, Coiflets, among others.)\footnote{
  This statement differs from the statement in the extremal
  setting that says that discarding all coefficients below $\tau$
  introduces $O(\tau\log n)$ error, since the latter does not account
  for the number of terms.}. We provide a $O(B+\log n)$ space and $O(n)$
  time one-pass algorithm in the data stream model.
  We give a modified greedy strategy,
  which is not normalization, but is similar to some scaling
  strategies used in practice. Our strategy demonstrates why several scaling
  based algorithms used in practice work well.

\item A surprising consequence of the above is an universal representation
using $O(B\log n)$ coefficients that {\em simultaneously} approximate the
signal for all $\ell_p$ distances up to $O(\log n)$.
 
\item The unrestricted optimization problem has a fully
  polynomial-time approximation scheme (FPTAS) for all $\ell_p$
  distances in the Haar system, that is, the algorithm runs in time
  polynomial in $B,\epsilon,n$.  The algorithm is one-pass,
  $n^{\frac1p}$ space and $n^{1+\frac1p}$ time for $\ell_p$
  distances. Therefore, the algorithm is a streaming algorithm with
  sublinear space for $p>1$.  For $\ell_\infty$, the algorithm runs in
  polylog space and linear time\footnote{For clarity here, we are
  suppressing terms based on $\log n$, $B$, and $\epsilon$. The exact
  statements appear in Theorems~\ref{mainthm} and \ref{mainthm2}.}.

\item For more general compactly supported systems we display how our
  ideas yield a quasi-polynomial time approximation scheme
  (QPTAS)\footnote{This implies that the running time is $2^{O(\log^c
  n)}$ for some constant $c$ ($c=1$ gives polynomial time).}.
\eat{These approximations show that the problem is unlikely to be {\sc
  NP--hard} in the strong sense and the difficulty in approximation is
  only precision based.}  
  This result is in contrast to the case of an arbitrary dictionary
  which, as we already mentioned, is hard to approximate to within any
  constant factor {\em even allowing} quasi-polynomial
  time\footnote{Follows from the result of Feige~\cite{F98}.}.

\item The results extend to fixed dimensions and workloads with increases in
  running time and space. 
\end{enumerate}

\item In terms of techniques, we introduce a new lower bounding
  technique using the basis vectors $\{\psi_i\}$, which gives us the
  above result regarding the gap between the restricted and
  unrestricted versions of the problem. We also show that bounds using
  the scaling vectors $\{\phi_i\}$ are useful for these optimization
  problems and, along with the lower bounds using $\{\psi_i\}$, give
  us the approximation schemes. To the best of our knowledge, this is
  the first use of both the scaling and basis vectors to achieve such
  guarantees.

\item We show that the lower bound for general compact systems can be
  extended to an approximation algorithm for adaptive quantization.
  This is the first approximation algorithm for this problem.

\item For tree-structured dictionaries composed of the type of
compactly supported wavelets we consider, our algorithms can be
combined with the dynamic programming algorithm of Coiffman and
Wickerhauser \cite{CW92} to find a $B$-term representation of the
given $f$.  The $\ell_p$ error of the representation we construct
provably approximates the error of a best representation of $f$ using
$B$ terms from a basis in the dictionary.

\end{enumerate}
\bigskip 
The key technique used in this paper is to lower bound the
solution based on a system of linear equations but with one non-linear
constraint. This lower bound is used to set the `scale' or `precision'
of the solution, and we show that the best solution respecting this
precision is a near optimal solution by `rounding' the components of 
the optimal solution to this precision. Finally the best solution in this class 
is found by a suitable dynamic program adapted to the data stream setting.

\bigskip
We believe that approximation algorithms give us the correct
standpoint for construction of approximate representations.  The
goal of approximation theory is to approximate representation; the
goal of approximation algorithms is to approximate optimization.  Data
stream algorithms are inherently approximate (and often randomized)
because the space restrictions force us to retain approximate
information about the input.  These goals, of the various uses of the
approximation, are ultimately convergent.

\bigskip
\paragraph*{Organization} We begin by reviewing some preliminaries
of wavelets.  In Section~\ref{sec:greedy} we present our greedy
approximation which also relates the restricted to the unrestricted
versions of the problem.  Section~\ref{sec:greedy_apps} presents 
applications of the greedy algorithm; namely, an approximate universal 
representation, approximation algorithms for adaptive quantization, 
and examples illustrating the use of non-$\ell_2$ norms for image 
representations.  Section~\ref{apxschemes} is the main section
of the paper wherein we present our approximation schemes.  We detail
the FPTAS for the Haar system and show its extensions to multiple
dimensions and workloads. We subsequently demonstrate in 
Section~\ref{sec:extensions} how the same ideas translate to a FPTAS 
for multi-dimensional signals and workloads, and a QPTAS under more 
general compactly supported wavelets. In Section~\ref{sec:best} we
present the tree-structured best-basis selection algorithm.  Finally,
in Section~\ref{sec:expt} we display some experimental results
contrasting the performance of an optimal algorithm that is restricted
to choosing Haar expansion coefficients with our Haar FPTAS.

\section{Preliminaries}\label{sec:prelim}

The problem on which we mainly concentrate is the following:
\bigskip

\begin{problem}[$B$-term Representation]\label{prob:intro:wavelets}
Given $f \in \real^n$, $p\in[1,\infty]$, a compactly-supported wavelet basis
for $\real^n$ $\{\psi_i\}_{i=1}^n$, and an integer $B$, find a solution 
$\{z_i\}_{i=1}^n$, $z_i\in\real$,  with at most $B$ non-zero components 
such that $\norm{f -\sum_i z_i\psi_i}_p$ is minimized.
\end{problem}

We will often refer to this problem as the \emph{unrestricted}
$B$-term representation problem in order to contrast it with a
\emph{restricted} version where the non-zero components of the
solution can only take on values from the set $\{\langle f,
\psi_i\rangle, i\in[n]\}$.  That is, in the restricted version, each
$z_i$ can only be set to a coefficient from the wavelet expansion of
$f$, or zero.

\subsection{Data Streams}

For the purpose of this paper, a data stream computation is a space
bounded algorithm, where the space is sublinear in the input.  Input
items are accessed sequentially and any item not explicitly stored
cannot be accessed again in the same pass.  In this paper we focus on
{\em one pass} data streams.  We will assume that we are given numbers
$f=f(1),\ldots,f(i),\ldots, f(n)$ which correspond to the signal $f$
to be summarized in the increasing order of $i$. This model is often
referred to as the \emph{aggregated model} and has been used widely
\cite{GMMO00,GKMS01,GIMS02}.  It is specially suited to model streams
of time series data \cite{keogh01,CKMP02} and is natural for
transcoding a single channel.  
\hide{Note that we can pretend that we know $n$ by losing a $O(\log n)$
factor in running time and space by running $O(\log n)$ algorithms in
parallel guessing $n$ to different powers of $2$.}%
Since we focus on dyadic wavelets (that are dilated by powers of $2$),
assuming $n$ is a power of $2$ will be convenient, but not necessary.
As is standard in literature on streaming~\cite{GKS01,GGIKMS02,I03},
we also assume that the numbers are polynomially bounded, i.e., all
$|f(i)|$'s are in the range $[n^{-c},n^c]$ for some constant $c$.
\eat{Our algorithms will depend on $\log M$ which is $ \sim \log
n$, and not on $M$.}

\subsection{Compactly Supported Wavelets} 

We include here some definitions and notation that we use in the main
text. Readers familiar with wavelets can easily skip this section.
For thorough expositions on wavelets, we refer the interested reader
to the authoritative texts by Daubechies~\cite{wave3} and
Mallat~\cite{wave2}. For a brief introduction to wavelets,
see~\cite[Chp.~2.3]{Harb07}.

A wavelet basis $\{\psi_k\}_{k=1}^n$ for $\real^n$ is a basis where
each vector is constructed by dilating and translating a single
function referred to as the \emph{mother wavelet} $\psi$.  For example
the Haar mother wavelet, due to Haar~\cite{Haar10}, is given by:
\[ \psi_H(t) = 
\left\{ \begin{array}{ll}
  \phantom{-}1 & \mbox{if } 0 \le t < 1/2 \\
  -1 & \mbox{if } 1/2 \le t < 1  \\
  \phantom{-}0 & \mbox{otherwise} 
\end{array}\right. \]
The Haar basis for $\real^n$ is composed of the vectors $\psi_{j,s}[i]
= 2^{-j/2}\psi_H\left(\frac{i - 2^j s}{2^j}\right)$ where $i\in [n]$,
$j = 1, \ldots, \log n $, and $s = 0, \ldots, n/2^j - 1$, plus their
orthonormal complement $\frac{1}{\sqrt{n}}1^n$.  This last basis
vector is closely related to the Haar \emph{multiresolution scaling
  function} $\phi_H(t) = 1$ if $0\le t < 1$ and $0$ otherwise.  In
fact, there is an explicit recipe for constructing the mother wavelet
function $\psi$ from $\phi$ using a \emph{conjugate mirror
  filter}~\cite{Mallat89,Meyer92} (see also Daubechies~\cite{wave2},
and Mallat~\cite{wave3}).  Notice that the Haar mother wavelet is
compactly supported on the interval $[0,1)$.  This wavelet, which was
discovered in 1910, was the only known wavelet of compact support
until Daubechies constructed a family of compactly-supported wavelet
bases~\cite{D88} in 1988 (see also~\cite[Chp.~6]{wave3}).

The vector $\psi_{j,s}$ is said to be centered at $2^js$ and of scale
$j$ and is defined on at most $ (2q-1)2^j - 2(q-1)$ points.  For ease
of notation, we will use both $\psi_i$ and $\psi_{j,s}$ depending on
the context and assume there is a consistent map between them.

\paragraph*{The Cascade Algorithm for computing 
$\langle f,\psi_{j,s}\rangle,\langle f,\phi_{j,s} \rangle$}

Assume that we have the conjugate mirror filter $h$ with support
$\{0,\ldots,2q-1\}$. Given a function $f\in\real^n$, we set
$a_0[i]=f[i]$, and repeatedly compute $a_{j+1}[t]= \sum_s
h[s-2t]a_j[s]$ and $ d_{j+1}[t]= \sum_s g[s-2t]a_j[s]$ (where $g[k] =
(-1)^k h[1-k]$ is also a conjugate mirror filter).  Notice that if the
filter $h$ has support $\{0,\cdots,2\taps-1\}$, then we have $0\leq
s-2t \leq 2\taps-1$.  This procedure gives $a_j[t] =\langle f,
\phi_{j,t} \rangle$ and $ d_j[t]=\langle f, \psi_{j,t} \rangle $.

In order to compute the inverse transform, we evaluate $a_{j}[t]=
\sum_s h[t-2s]a_{j+1}[s] + \sum_s g[t-2s]d_{j+1}[s] $.  Observe that
by setting a single $a_j[s]$ or $d_{j}[s]$ to $1$ and the rest to $0$,
the inverse transform gives us $\phi_{j,s}$ or $\psi_{j,s}$. Indeed,
this is the algorithm usually used to compute $\phi_{j,s}$ and
$\psi_{j,s}$.  \smallskip

We will utilize the following proposition which is a consequence of
the dyadic structure of compactly-supported wavelet bases.

\begin{proposition}\label{prop:qlogn-basis}
  A compactly-supported wavelet whose filter has $2\taps$ non-zero
  coefficients generates a basis for $\real^n$ that has $O(\taps \log
  n)$ basis vectors with a non-zero value at any point $i\in[n]$.
\end{proposition}

\section{Greedy Approximation Algorithms for General Compact Systems and
Data Streams}
\label{sec:greedy}

Recall our optimization problem: Given a compactly-supported wavelet basis
$\{\psi_i\}$ and a target vector $f$, we wish
to find $\{z_i\}$ with at most $B$ non-zero numbers to minimize $\| f
- \sum_i z_i \psi_i \|_p$.

We present two analyses below corresponding to $\ell_\infty$
and $\ell_p$ errors when $p\in [1,\infty)$.  In each case we begin by
analyzing the sufficient conditions that guarantee the error.  
A (modified) greedy coefficient retention algorithm will naturally
fall out of both analyses. The proof shows that several of the
algorithms that are used in practice have bounded approximation
guarantee. Note that the optimum solution can choose any values in the
representation $\hat{f}$.

In what follows the pair $(p,\q)$ are the usual conjugates; i.e.,
$\frac1p+\frac1\q = 1$ when $1 < p < \infty$, and when $p = 1$ we
simply set $\q=\infty$.  For simplicity, we start with the $p=\infty$
case.

\subsubsection{An $\ell_\infty$ Algorithm and Analysis}
\label{sec:greedyInf}

The main lemma, which gives us a lower bound on the optimal error, is:
\begin{lemma}
\label{lb}
  Let $\E$ be the minimum error under the $\ell_\infty$ norm and
$\{z_i^*\}$ be the
  optimal solution, then
  \[ - \| \psi_i \|_1 |\E| \leq \langle f, \psi_i \rangle - z^*_i \leq \|
\psi_i \|_1 |\E| \enspace .\]
\end{lemma}
\begin{proof}
For all $j$ we have $ -|\E| \leq f(j) - \sum_i z^*_i
\psi_i(j) \leq |\E|$.  Since the equation is symmetric multiplying it
by $\psi_k(j)$ we get,
\[ -|\E| |\psi_k(j)|\ \leq
f(j)\psi_k(j) - \psi_k(j) \sum_i z^*_i \psi_i(j)\ \leq
|\E| |\psi_k(j)| \]
If we add the above equation for all $j$, since $-|\E| \sum_j
|\psi_k(j)| = -|\E| \|\psi_k\|_1$ we obtain (consider only the left
side)
\begin{align*}
-|\E| \|\psi_k\|_1
& \leq \sum_j f(j)\psi_k(j) - \sum_j \psi_k(j) \sum_i z^*_i  \psi_i(j) \\
& =  \langle f, \psi_k \rangle -  \sum_i z^*_i \sum_j \psi_k(j) \psi_i(j)
\\
& = \langle f, \psi_k \rangle -  \sum_i z^*_i \delta_{ik} = \langle f,
\psi_k \rangle - z^*_k
\enspace .
\end{align*}
The upper bound follows analogously.
\end{proof}

\paragraph*{A Relaxation.} Consider the following program:
\begin{eqnarray}
& & \hspace{-1.7in} \text{minimize } \tau \nonumber \\
-\tau \|\psi_1\|_1 & \leq \langle f, \psi_1 \rangle - z_1 & \leq \tau
\|\psi_1\|_1 \nonumber \\
\vdots & \vdots & \vdots \label{sys} \\
-\tau \|\psi_n\|_1 & \leq \langle f, \psi_n \rangle - z_n & \leq \tau
\|\psi_n\|_1 \nonumber \\
& & \hspace{-1.7in} \mbox{ At most $B$ of the $z_i$'s are non-zero }
\nonumber
\end{eqnarray}
\noindent Observe that $\E$ is a feasible solution for the above
program and $\E \geq \tau^*$ where $\tau^*$ is the optimum value of the program.  
Also, Lemma \ref{lb} is not specific to
wavelet bases, and indeed we have $\E = \tau^*$ when $\{\psi_i\}$ is the
standard basis, i.e.~$\psi_i$ is the vector with $1$ in the
$i^\text{th}$ coordinate and 0 elsewhere. The next lemma is
straightforward.
\begin{lemma}
\label{first}
The minimum $\tau$ of program \eqref{sys} is the $(B+1)^{th}$ largest
value $\frac{\abs{\langle f, \psi_i \rangle}}{\|\psi_i\|_1}$.
\end{lemma}

\paragraph*{The Algorithm.} We choose the largest $B$
coefficients based on $|\langle f, \psi_i \rangle|/\|\psi_i\|_1$. This
can be done over a one pass stream, and in $O(B+\log n)$ space for any
compact wavelet basis.
Note that we  need not choose $z_{i}
= \langle f, \psi_{i} \rangle$ but any $z_i$ such that $|z_{i} -
\langle f, \psi_{i} \rangle|/\|\psi_{i}\|_1 \leq \tau^*$.  But in
particular, we may choose to retain coefficients and set $z_{i} =
\langle f, \psi_{i} \rangle$. The alternate choices may (and often
will) be better.
Also note that the above is only a necessary condition;
we {\em still} need to analyze the guarantee provided by the
algorithm.

\begin{lemma}
\label{second}
  For all basis vectors $\psi_i$ of a compact system
%%%%%computed by the cascade algorithm -- why are we saying this ?
  there exists a constant $C$ s.t.
  $\|\psi_i\|_p \|\psi_i\|_\q \leq \sqrt{\taps}C$.
\end{lemma}
\begin{proof}
  Suppose first that $p < 2$.  Consider a basis vector $\psi_i[] =
  \psi_{j,s}[]$ of sufficiently large scale $j$ that has converged to
  within a constant $r$ (point-wise) of its continuous analog
  $\psi_{j,s}()$~\cite[pp.~264--5]{wave2}.  That is,
  $\abs{\psi_{j,s}[k] - \psi_{j,s}(k)} \le r$ for all $k$ such that
  $\psi_{j,s}[k]\ne 0$. The continuous function $\psi_{j,s}()$ is
  given by $\psi_{j,s}(t) = 2^{-j/2}\psi(2^{-j}t -s)$, which implies
  $\psi_{j,s}[k] = O\left(2^{-j/2}\psi(2^{-j}k - s)\right) =
  O(2^{-j/2})$.  Note that we are assuming $\norm{\psi}_\infty$ itself
  is some constant since it is independent of $n$ and $B$. Combining
  the above with the fact that $\psi_{j,s}[]$ has at most $(2q)2^j$
  non-zero coefficients, we have
  $\norm{\psi_{j,s}}_\q=O(2^{-j/2}((2\taps)2^j)^{1/\q}) =
  O(2^{j(\frac1\q-\frac12)}(2\taps)^{\frac1\q})$.%
\hide{
Suppose first that $p < 2$. Consider a basis vector
$\psi_i=\psi_{j,s}$ of sufficiently large (constant) scale $j$ such
that $2^{(j-1)/2}\phi_{j-1,s}$ has converged to within a constant $r$
(point-wise) of $\phi$. Note that we are assuming $\|\phi\|_\infty$
itself is some constant since it is independent of $n$ and $B$. Since
$\psi_{j,s}$ is defined in terms of $\phi_{j-1,s}$ and has at most
$(2q)2^j$ non-zero coefficients, we have
$\|\psi_{j,s}\|_\q=O(2^{-j/2}((2\taps)2^j)^{1/\q}) 
= O(2^{j(\frac1\q-\frac12)}(2\taps)^{\frac1\q})$.
%%% Note: $\|\psi_i\|_\infty$ is at most $(\|\psi\|_\infty+r) 2^{-j/2}$.
}

Now by H\"{o}lder's inequality, $\norm{\psi_{j,s}}_p \leq ((2
\taps)2^j)^{\frac1p -\frac12}\|\psi_{j,s}\|_2 =
2^{j(\frac1p-\frac12)}(2 \taps)^{\frac1p-\frac12}$.  Therefore, for
sufficiently large scales $j$,
$\norm{\psi_{j,s}}_p\norm{\psi_{j,s}}_\q =
O(2^{j(\frac1p+\frac1\q-1)}(2 \taps)^{\frac1p+\frac1\q-\frac12}) =
O(\sqrt{\taps})$, and the lemma holds. For basis vectors at smaller
(constant) scales, since the number of non-zero entries is constant,
the $\ell_p$ norm and the $\ell_\q$ norm are both constant.

Finally, for $p > 2$, the argument holds by symmetry.
\end{proof}

\begin{theorem}
The $\ell_\infty$ error of the final approximation is at most
$O(\taps^{3/2}\log n)$ times $\E$ for any compactly supported wavelet.
\label{greedyLinf}
\end{theorem}
\begin{proof}
  Let $\{z_i\}$ be the solution of the system \eqref{sys}, and let the
  set of the inner products chosen be ${\cal S}$.  Let $\tau^*$ is the
  minimum solution of the system \eqref{sys}. The $\ell_\infty$ error
  seen at a point $j$ is $|\sum_{i \not \in {\cal S}} \langle f,
  \psi_i \rangle \psi_i(j)| \le \sum_{i\not \in {\cal S}} |\langle f,
  \psi_i \rangle||\psi_i(j)|$.  By Lemma~\ref{first}, this sum is at
  most $\sum_{i\not \in {\cal S}} \tau^*\|\psi_i\|_1|\psi_i(j)|$,
  which is at most
  $\tau^*\max_{i\not\in\mathcal{S}}\|\psi_i\|_1\|\psi_i\|_\infty$
  times the number of vectors that are non-zero at $j$.  By
  Proposition~\ref{prop:qlogn-basis} the number of non-zero vectors at
  $j$ is $O(\taps \log n)$.  By Lemma~\ref{second},
  $\|\psi_i\|_1\|\psi_i\|_\infty \le \sqrt{\taps}C$ for all $i$, and
  since $\tau^*\le \E$ we have that the $\ell_\infty$ error is bounded
  by $O(\taps^{3/2}\log n)\E$.
\end{proof}

\subsubsection{An $\ell_p$ Algorithm and Analysis for $p\in[1,\infty)$}
\label{sec:greedyLp}

Under the $\ell_p$ norm, a slight modification to the algorithm above
also gives an $O(\taps^{3/2}\log n)$ approximation guarantee.

\begin{lemma}
\label{lb2}
  Let $\E$ be the minimum error under the $\ell_p$ norm and $\{z_i^*\}$ be
the
  optimal solution, then for some constant $c_0$,
  \[ \left(\sum_k \frac{1}{\|\psi_k\|_\q^p} |\langle f, \psi_k \rangle -
  z^*_k|^p\right)^{\frac1p} \leq \left(c_0 \taps \log n\right)^{\frac1p}\E \enspace .\]
\end{lemma}
\begin{proof}
 An argument similar to that of Lemma~\ref{lb} gives

\begin{IEEEeqnarray*}{lrCl}
& \sum_i\left|f_i\psi_k(i) - \sum\nolimits_j
z^*_j\psi_j(i)\psi_k(i)\right|
  &\ =\ & \sum_i \xi_i\abs{\psi_k(i)} \leq 
\left(\sum_{i \in \text{ support of $\psi_k$}} \xi_i^p\right)^{1/p} \|\psi_k\|_\q \\
\Rightarrow  & \frac{1}{\|\psi_k\|_\q^p} \abs{\langle f, \psi_k\rangle - z_k^*}^p
  &\ \le\ &  \sum_{i \in \text{ support of $\psi_k$}} \xi_i^p \\
\Rightarrow & \sum_k \frac{1}{\|\psi_k\|_\q^p} \abs{\langle f,\psi_k\rangle - z_k^*}^p
  &\ \le\ & c_0 \taps \log n \sum_i \xi_i^p \enspace ,
\end{IEEEeqnarray*}

where the last inequality follows from
Proposition~\ref{prop:qlogn-basis}, that each $i$ belongs to
$O(\taps\log n)$ basis vectors ($c_0$ is the constant hidden by the
this $O$-term).
\end{proof}

\paragraph*{A Relaxation.} Consider the following system of equations,
\begin{eqnarray}
& & \hspace{-2.58in} \text{minimize } \tau \nonumber \\
\left(\sum_{i=1}^n 
  \frac{\abs{\langle f, \psi_i\rangle - z_i}^p}{\|\psi_i\|_\q^p} 
\right)^{\frac1p}
& \le\ (c_0 \taps\log n)^{\frac1p}\tau  \label{sysL1}\\
& & \hspace{-2.62in} \mbox{ At most $B$ of the $z_i$'s are non-zero }
\nonumber
\end{eqnarray}

\paragraph*{The Algorithm.} We choose the largest $B$
coefficients based on $|\langle f, \psi_k \rangle|/\|\psi_k\|_\q$,
which minimizes the system~\eqref{sysL1}. This computation
can be done over a one pass stream, and in $O(B+\log n)$ space.

\begin{theorem}
\label{greedyL1}
Choosing the $B$ coefficients $\langle f, \psi_k\rangle$ that are
largest based on the ordering $\abs{\langle f, \psi_k\rangle}/\|\psi_k\|_\q$ is
a streaming $O(\taps^{3/2}\log{n})$ approximation algorithm for the
unrestricted optimization problem under the $\ell_p$ norm.
\end{theorem}
Note this matches the $\ell_\infty$ bounds, but
stores a (possibly) different set of coefficients.

\begin{proof}
  Let the value of the minimum solution to the above system of
  equations~\eqref{sysL1} be $\tau^*$. Since $\{z_i^*\}$ is feasible
  for system~\eqref{sysL1}, $\tau^*\leq \E$.  Assume $\mathcal{S}$ is the
  set of coefficients chosen, the resulting error $\E_\mathcal{S}$ is,
\begin{IEEEeqnarray*}{rCl}
\E_\mathcal{S}^p &\ =\ &
  \sum_i\left|\sum_{k\not\in\mathcal{S}}\langle f,
\psi_k\rangle\psi_k(i)\right|^p
\le\
  \sum_i (c_0 \taps\log n)^{p-1}\sum_{k\not\in\mathcal{S}}
    \abs{\langle f,\psi_k\rangle}^p\abs{\psi_k(i)}^p \\
& = & 
  (c_0 \taps\log n)^{p-1} 
  \sum_{k\not\in\mathcal{S}} 
    \abs{\langle f, \psi_k\rangle}^p\norm{\psi_k}_p^p \\
& \le & 
  (c_0 \taps\log n)^{p-1} 
  \sum_{k\not\in\mathcal{S}} \frac{C^p \taps^{\frac{p}{2}}}{\|\psi_k\|_\q^p}
    \abs{\langle f, \psi_k\rangle}^p \\
& = &
   C^p \taps^{\frac{p}{2}} (\tau^* c_0 \taps \log n)^p \enspace .
\end{IEEEeqnarray*}
Here, the first inequality is H\"older's inequality combined with
Proposition~\ref{prop:qlogn-basis} and the fact that $p/\q = p-1$; the
second inequality follows from Lemma~\ref{second}; and the final
equality follows from the optimality of our choice of coefficients for
the system~\eqref{sysL1}.  Now since $\tau^* \le \E$, we have that 
$\E_\mathcal{S} \le c_0 C \taps^{\frac32} \E \log n$.
\end{proof}

\subsubsection{Summary and a Tight Example}
In the two preceeding subsections we showed the following:
\begin{theorem}\label{greedyLp3}
  Let $\frac1p + \frac1\q = 1$.  Choosing the largest $B$ coefficients
  based on the ordering $|\langle f, \psi_i\rangle|/\|\psi_i\|_\q$,
  which is possible by a streaming $O(B+\log n)$ algorithm, gives a
  $O(\taps^{\frac32} \log n)$ approximation algorithm for the
  unrestricted optimization problem
  (Problem~\ref{prob:intro:wavelets}) under the given $\ell_p$
  norm. The argument naturally extends to multiple dimensions.
\end{theorem}

As is well-known, this choice of coefficients is optimal when $p=2$
(since $\q=2$ and $\|\psi_i\|_2=1$). 

\emph{Note that the above theorem bounds the gap between the
restricted (where we can only choose wavelet coefficients of the input in
the representation) and unrestricted optimizations.}

\paragraph*{A tight example for the $\ell_\infty$ measure.}
Suppose we are given the Haar basis $\{\psi_i\}$ and the vector $f$
with the top coefficient $\langle f, \psi_1\rangle = 0$ and with
$\langle f, \psi_i\rangle/\norm{\psi_i}_1 = 1-\epsilon$ for $i \le
n/2$, and $\langle f, \psi_i\rangle/\norm{\psi_i}_1 = 1$ for $i > n/2$
(where $\psi_i$, $i > n/2$, are the basis with smallest support).  Let
$B = n/c - 1$ where $c \ge 2$ is a constant that is a power of $2$.
The optimal solution can choose the $B$ coefficients which are in the
top $\log n - \log c$ levels resulting in an error bounded by $\log
c$. The $\ell_\infty$ error of the greedy strategy on the other hand
will be at least $\log n -1$ because it will store coefficients only
at the bottom of the tree.  Hence it's error is at least $\log n/\log
c - o(1)$ of the optimal.

% End hide 

\section{Applications of the Greedy Algorithm}\label{sec:greedy_apps}

Our greedy algorithm extends to a variety of scenarios, which
illustrate the scope and the applicability of the techniques presented
above.

\subsection{A Universal Representation}
\label{sec:universal}

In this section we present a strategy that stores $B(\log n)^2$
coefficients and simultaneously approximates the optimal
representations for all $p$-norms.  Notice that in
Problem~\ref{prob:intro:wavelets} we know the $p$-norm we are trying
to approximate.  Here, we do \emph{not} know $p$ and we wish to come
up with a representation such that for all $p\in [1,\infty]$, its
error measured with $\snorm{f - \hat{f}_u}_p$ is $O(\log n)$ times the
optimal error $\min_z \norm{f - \sum_i z_i\psi_i}_p$ where $x$ has at
most $B$ non-zero components.  Notice that we allow our universal
representation to store a factor $(\log n)^2$ more components than any
one optimal representation; however, it has to approximate all of them
concurrently.

We run our algorithm as before computing the wavelet coefficients of
the target vector $f$; however, we need to determine which
coefficients to store for our universal representation.  To this end,
define the set:
\begin{equation}\label{eq:pnormset}
{\cal N} = \{p_t : p_t = 1+\frac{t}{\log n},\; t = 0,\ldots, \log n(\log n -1) \} \enspace . 
\end{equation}
For every $p_t \in {\cal N}$, we will store the $B$ coefficients that
are largest based on the ordering $\abs{\langle f,
  \psi_k\rangle}/\|\psi_k\|_{\q_t}$ where $\q_t$ is the dual norm to
$p_t$.  Hence, the number of coefficients we store is no more than
$B(\log n)^2$ since $\abs{{\cal N}} = (\log n)^2$.  Note that our dual
programs show that for a given $p$, storing more than $B$ coefficients
does not increase the error of the representation. Now let $\hat{f}_u$
be our resultant representation; i.e., if ${\cal S}$ contains the
coefficients we chose, then $\hat{f}_u = \sum_{i\in {\cal S}} \langle
f, \psi_i \rangle \psi_i$; and let $f^*_{(p)}$ be the optimal
representation under the norm $\ell_p$.  Consider first the case when
$p \in (p_t,\, p_{t+1})$ where $p_t,p_{t+1} \in{\cal N}$.
\begin{IEEEeqnarray}{rCll}
\snorm{f - \hat{f}_u}_p & \le & \snorm{f - \hat{f}_u}_{p_t} 
  & \mbox{since } p > p_t \nonumber\\
& \le & cq^\frac32(\log n)\snorm{f - f^*_{(p_t)}}_{p_t} 
  & \mbox{by Thereom~\ref{greedyLp3}} \nonumber\\
& \le & cq^\frac32(\log n)\snorm{f - f^*_{(p)}}_{p_t} 
  & \mbox{by the optimality of } f^*_{p_t} \mbox{ for } \ell_{p_t} \nonumber\\
& \le & cq^\frac32(\log n) n^{\frac{1}{p_t} - \frac{1}{p}}\snorm{f - f^*_{(p)}}_{p} \quad 
  & \mbox{by H\"older's inequality}\label{eq:universal}
\end{IEEEeqnarray}
However $1/p_t - 1/p \le 1/p_t - 1/p_{t+1}$ since $p < p_{t+1}$; 
and by their definition,
\[ \frac{1}{p_t} - \frac{1}{p_{t+1}} = 
\frac{\log n}{(\log n + t)(\log n + t + 1)} \le 
\frac{1}{\log n} \enspace . \]
Hence, $n^{\frac{1}{p_t}- \frac{1}{p}} \le n^{1/(\log n)} = 2$; and
from expression~\eqref{eq:universal} we have that $\snorm{f -
  \hat{f}_u}_p = O(q^\frac32 \log n) \snorm{f - f^*_{(p)}}_{p}$ as
required.  When $p > p_{t}$ for $t = \log n(\log n -1)$, we
immediately have $n^{\frac{1}{p_t}- \frac{1}{p}} \le n^{1/(\log n)}$
and the result follows.

\subsection{Adaptive Quantization}\label{sec:adaptive}
Wavelets are extensively used in the compression of images and audio
signals. In these applications a small percent saving of space is
considered important and attention is paid to the bits being stored.
The techniques employed are heavily engineered and typically designed
by some domain expert. The complexity is usually two-fold: First, the
numbers $z_i$ do not all cost the same to represent. In some
strategies; e.g., strategies used for audio signals, the number of
bits of precision to represent a coefficient $z_i$ corresponding to
the basis vector $\psi_i=\psi_{j,s}$ is fixed, and it typically
depends only on the scale $j$. (Recall that there is a mapping from
$\psi_i$ to $\psi_{j,s}$.) Further the $a_j[]$'s are computed with a
higher precision than the $d_j[]$'s. This affects the space needed by
the top-most coefficients. In yet another strategy, which is standard
to a broad compression literature, it is assumed that $\log_2 z$ bits
are required to represent a number $z$. All of these bit-counting
techniques need to assume that the signal is bounded and there is some
reference unit of precision.

Second, in several systems, e.g., in
JPEG2000~\cite{christopoulos2000jsi}, a bitmap is used to indicate the
non-zero entries. However the bitmap requires $O(n)$ space and it is
often preferred that we store only the status of the non-zero values
instead of the status of all values in the transform. In a setting
where we are restricted to $o(n)$ space, as in the streaming setting,
the space efficiency of the map between non-zero coefficients and
locations becomes important. For example, we can represent $\psi_i =
\psi_{j,s}$ using $\log\log n + \log(n/2^j)+O(1)$ bits instead of
$\log n$ bits to specify $i$. Supposing that only the vectors with
support of $\sqrt{n}$ or larger are important for a particular signal,
we will then end up using half the number of bits. Notice that this
encoding method {\em increases} the number of bits required for
storing a coefficient at a small scale $j$ to more than $\log n$. This
increase is (hopefully) mitigated by savings at larger scales. Note
also that the wavelet coefficients at the same level are treated
similarly.

The techniques we presented in Section~\ref{sec:greedy} naturally
extend to these variants of the bit-budget problem. In what follows,
we consider three specific cases:
\begin{enumerate}
\item {\em Spectrum Representations:} The cost $c_i$ of storing a
  coefficient corresponding to $i$ is fixed. This case includes the
  suggested strategy of using $\log\log n + \log(n/2^j)+O(1)$ bits.

\item {\em Bit Complexity Representations:} The cost of storing the
  $i$th coefficient with value $z_i$ is $c_i + b(z_i)$ for some
  (concave) function $b()$. A natural candidate for $b()$ is
  $b(z)=O(1) - \log z_i^{\text{frac}}$ where $z_i^{\text{frac}}$ is
  the fractional part of $z_i$ and is less than $1$ (thus $-\log
  z_i^{\text{frac}}$ is positive).  This encodes the idea that we can
  store a higher ``resolution'' at a greater cost.

\item {\em Multiplane Representations:} Here the data conceptually
  consists of several ``planes'', and the cost of storing the $i$th
  coefficient in one plane depends on whether the $i$th coefficient in
  another plane is retained.  For example, suppose we are trying to
  represent a RGBA image which has four attributes per pixel.  Instead
  of regarding the data as $4\times 2$ dimensional, it may be more
  useful, for example if the variations in color are non-uniform, to
  treat the data as being composed of several separate planes, and to
  construct an optimization that allocates the bits across them.

\end{enumerate}

The fundamental method by which we obtain our approximate solutions to
the above three problems is to use a greedy rule to lower bound the
errors of the optimal solutions using systems of constraints as we did
in Section~\ref{sec:greedy}.  We focus only on the $\ell_\infty$ error
for ease of presentation. As before, the techniques we use imply
analogous results for $\ell_p$ norms.

\subsubsection{Spectrum Representations}
In the case where the cost of storing a number for $i$ is a fixed
quantity $c_i$ we obtain a lower bound via a quadratic program that is
similar to \eqref{sys} using Lemma~\ref{lb}.  That is, $\text{minimize
} \tau$ with the constraints $x_i \in \{0,1\} $ and $\sum_i x_i c_i
\leq B$, and for all $i$
\begin{equation}\label{sys2}
-\tau \|\psi_i\|_1\ \leq\ \langle f, \psi_i \rangle - x_i z_i\ \leq\ \tau \|\psi_i\|_1  
\end{equation}
The program above can be solved optimally since the $c_i$'s are
polynomially bounded\hide{; however, the solution does not stream}. We
sort the coefficients in non-increasing order of $y_i := \abs{\langle
  f, \psi_i\rangle}/\norm{\psi_i}_1$.  If $y_{i_1} \ge y_{i_2} \ge
\cdots \ge y_{i_n}$, then we include coefficients $i_1, \ldots , i_k$
where $\sum_{j=1}^k c_{i_j} \le B < \sum_{j=1}^{k+1} c_{i_j}$.  The
value $y_{i_{k+1}}$ is then a lower bound on the error $\E$ of the
optimal representation $z^*$. Note that $z^*$ is a feasible solution
to program~\eqref{sys2}.  Hence, either $z^*$ includes coefficients
$i_1, \ldots, i_k$ in which case it cannot choose coefficient
$i_{k+1}$ for it will exceed the space bound $B$, and we have that $\E
\ge y_{i_{k+1}}$ (the optimal does not necessarily set $z_i =
\abs{\langle f, \psi_{i}\rangle}$); or, $z^*$ does not include one of
$i_1, \ldots, i_k$, thus $\E$ is again greater then or equal to
$y_{i_{k+1}}$. A proof similar to that of Theorem~\ref{greedyLinf}
shows that the error of our solution is $O(\log n)\E$.  \hide{Finally,
  our solution can be made to stream by means of a guessing strategy
  similar to that employed in Section~\ref{sec:bitcomplexity} below
  (where the cost function $b() = 0$).}

%% Keeping the hidden text below for later reference.
\hide{
however, the solution does not stream.  We can
also exceed space bounds by at most one item and solve the above by
sorting the coefficients in non-increasing order of $|\langle f,
\psi_i\rangle|/\|\psi_i\|_1 $. We include coefficients in an
hypothetical solution in this order. (Recall, coefficient $i$ costs
$c_i$.) We stop when we exceed $B$ (say at $B'$) and the next unchosen
coefficient is a lower bound for the largest (suitably weighted)
coefficient that is not chosen by the optimum solution. Thus we have a
lower bound on the error.

However, one issue needs to be addressed.  The greedy solution (for
any compact wavelet) has space $B'$. The dynamic program for the Haar
case should also be run for this value $B'$---for otherwise we cannot
argue that there is a feasible solution.  Thus in summary, in both
cases we achieve a {\em bi-criteria} approximation where we find a
solution whose $\ell_\infty$ error is $(1+\epsilon)$ times the error
of the optimum $B$-bit solution; however, our solution is allowed to
exceed $B$ by the space required to store an extra coefficient, which
is at most $\max_i c_i$.

For data streams in the Haar case, we cannot sort the coefficients, but
we can guess the value of the largest unchosen coefficient and perform the
dynamic program for $B+\max_i c_i$ for each guess.
} %% End hide.

\subsubsection{Bit Complexity Representations}\label{sec:bitcomplexity}
In the case where the cost is dependent on $z_i$ we cannot write an
explicit system of equations as we did in the case of spectrum
representations. However, we can guess $\tau$ up to a factor of $2$
and verify if the guess is correct.

In order to verify the guess, we need to be able to solve equations of
the form $\min_z b(z)$ s.t.~$|a-z|\leq t$ (since this is the format of
our constraints).  This minimization is solvable for most reasonable
cost models; e.g., if $b(z)$ is monotonically increasing.  As the
coefficients are generated, we compute $c_i + b(z_i)$ if $z_i\neq 0$,
where $z_i={\rm argmin}_z b(z) \mbox{ s.t. }|\langle f,\psi_i\rangle -
z|\leq t\norm{\psi_i}_1$ for our guess $t$ of the error.  If we exceed
the alloted space $B$ at any point during the computation, we know
that our guess $t$ is too small, and we start the execution over with
the guess $2t$.  Note that the optimal representation is a feasible
solution with value $\E$ and bit complexity $B$. Applying the analysis
of Section~\ref{sec:greedyInf} shows that the first solution we obtain
that respects our guess is a $O(\log n)$ approximation to the optimal
representation.

Since we assume that the error $\E$ is polynomially bounded, the above
strategy can be made to stream by running $O(\log n)$ greedy
algorithms in parallel each with a different guess of $\tau$ as above.

%% This is not necessary, but keeping it for later reference.
\hide{In this case, if we exceed the space $2B$ at any point, we know that our guess
$t$ is incorrect.  We may actually need space $2B$ because when we
exceed $B$ we may exceed it by $b(z)$ which is a large
quantity. However, $b(z)$ has to be less than $B$ for any solution and
this limits the space to $2B$. In this case we get a bi-criteria
approximation with a worst case blowup of factor $2$ in space usage.
} %% End hide.

\subsubsection{Multiplane Representations}
In this case we are seeking to represent data that is conceptually in
several ``planes'' simultaneously; e.g., RGBA in images. We could also
conceptualize images of the same object at various frequencies or
technologies.  The goal of the optimization is to allocate the bits
across them.  However, notice that if we choose the $i$th coefficient
for say the Red and the Blue planes (assuming that we are indicating
the presence or absence of a coefficient explicitly which is the case
for a sparse representation), then we can save space by storing the
fact that ``coefficient $i$ is chosen'' only once. This is easily
achieved by keeping a vector of four bits corresponding to each chosen
coefficient.  The values of the entries in the bit vector inform us if
the respective coefficient value is present. Therefore, the bit vector
$1010$ would indicate that the next two values in the data correspond
to Red and Blue values of a chosen coefficient.  Similarly, a vector
$1011$ would suggest that three values corresponding to Red, Blue and
Alpha are to be expected.

In what follows, we assume that the data is $D$ dimensional and it is
comprised of $t$ planes (in the RGBA example $D=2$ and $t=4$).  We are
constrained to storing at most $B$ bits total for the bit vectors, the
indices of the chosen coefficients, and the values of these
coefficients. For simplicity we assume that we are using the
$\ell_\infty$ error across all the planes. Otherwise, we would also
have to consider how the errors across the different planes are
combined.

We construct our approximate solution by first sorting the
coefficients of the $t$~planes in a single non-increasing order while
keeping track of the plane to which each coefficient belongs. As
before, we add the coefficients that are largest in this ordering to
our solution, and stop immediately before the coefficient whose
addition results in exceeding the alloted space $B$.  Note that if we
had added the $i$th coefficient of the Red plane first, and thereafter
wanted to include the Blue plane's $i$th coefficient, then we need
only account for the space of storing the index $i$ and the associated
bit vector when we add the coefficient for the first (in this case
Red) plane. The subsequent $i$th coefficients only contribute to the
cost of storing their values to the solution. (We can think of the
cost of storing each coefficient as fixed \emph{after} the ordering of
the coefficients is determined.) This strategy is reminiscent of the
strategy used by Guha, Kim and Shim \cite{GKS04} to lower bound the
optimum error for a similar problem in the $\ell_2$ setting.

The first coefficient we did not choose using this greedy selection
process is a lower bound on the optimal representation error.  Now, an
argument similar to that of Theorem~\ref{greedyLinf} shows that the
error of the resulting solution is a $O(\log n)$ factor away from the
error of the optimal solution.

%% End hide.

\subsection{Sparse Image Representation under non-$\ell_2$ Error Measures}  
\label{sec:imagecompress}

In this section we give three examples that demonstrate uses for our greedy algorithm in
compressing images.  A non-streaming version of the algorithm for Haar
and Daubechies wavelets was implemented in {\sc Matlab} using the {\it
  Uvi\_Wave.300} toolbox\footnote{For compatibility with our version
  of {\sc Matlab}, slight modifications on the toolbox were
  performed. The toolbox can be obtained from
  \url{http://www.gts.tsc.uvigo.es/~wavelets/}.}~\cite{uviwave}.
Pseudocode of the implementation is provided below in
Figure~\ref{fig:daubgreedy}.  The algorithm takes four parameters as
input: the image $X$, the number of coefficients to retain $B$, the
$p$-norm to minimize, and the type of Daubechies wavelet to use.  The
last parameter, $\taps$, determines the number of non-zero
coefficients in the wavelet filter.  Recall that the Haar wavelet is
the Daubechies wavelet with smallest support; i.e., it has $\taps =
1$.

\begin{figure}[ht]
\framebox[5.94in]{\parbox{5.65in}{
\begin{algorithm}{DaubGreedy}[X,B,p,\taps]{\label{alg:daubgreedy}}
\qcom{$X$ is a grayscale image (intensity matrix)}\\
Perform a 2D wavelet transform of $X$ using the Daubechies $D_q$ wavelet\\
Let $w$ be the wavelet coefficients of the transform\\
$\q \qlet p/(p-1)$\\
$y_i \qlet \abs{w_i}/\norm{\psi_i}_\q$\\
Let $\mathcal{I}$ be the indices of the $B$ largest $y_i$'s\\
$w_i \qlet 0$ if $i\not\in\mathcal{I}$\\
Perform a 2D inverse wavelet transform on the resulting $w$\\
Let $X'$ be the resulting image representation\\
\qreturn $X'$
\end{algorithm}
}}
\caption[Pseudocode of the greedy algorithm's implementation]
{Pseudocode of the greedy algorithm's implementation.}
\label{fig:daubgreedy}
\end{figure}

\begin{figure}
\centering
\subfigure[The original image]{\label{fig:originalpoem}
\begin{minipage}[t]{2.83in}
\centering \includegraphics[width=2.83in]{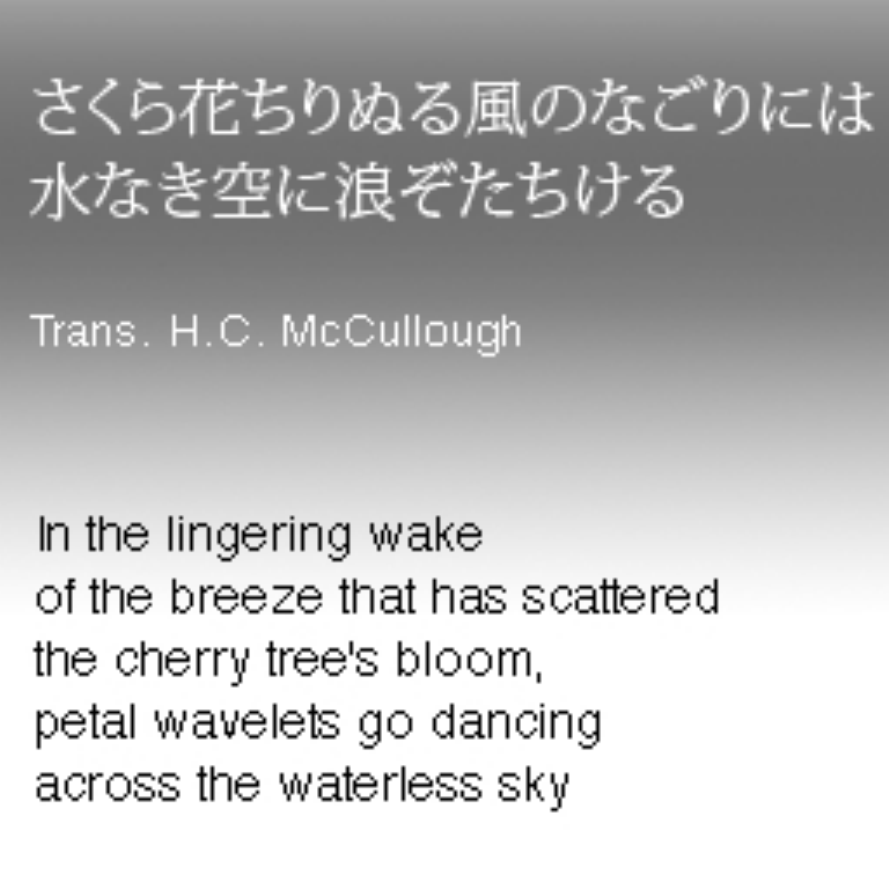}
\end{minipage}
} \subfigure[Output of the optimal $\ell_2$ algorithm 
(which retains the largest $B$ wavelet coefficients)]{\label{fig:L2poem}
\begin{minipage}[t]{2.83in}
\centering \includegraphics[width=2.83in]{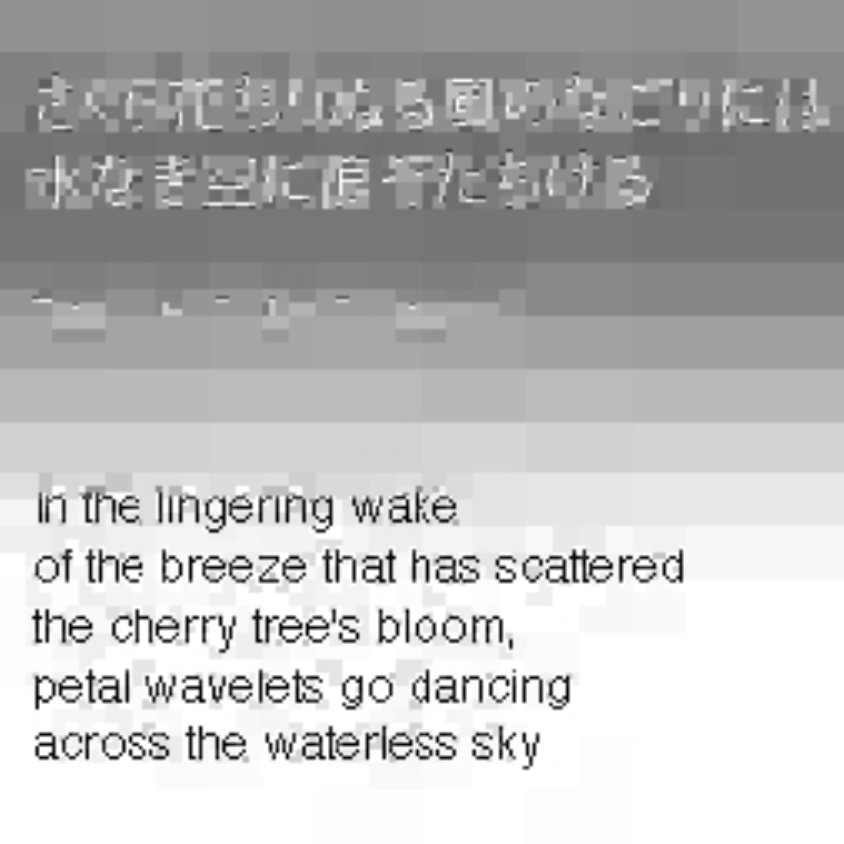}
\end{minipage}
} \\ \subfigure[Output of our greedy algorithm under $\ell_\infty$]{\label{fig:LInfpoem}
\begin{minipage}[t]{2.83in}
\centering \includegraphics[width=2.83in]{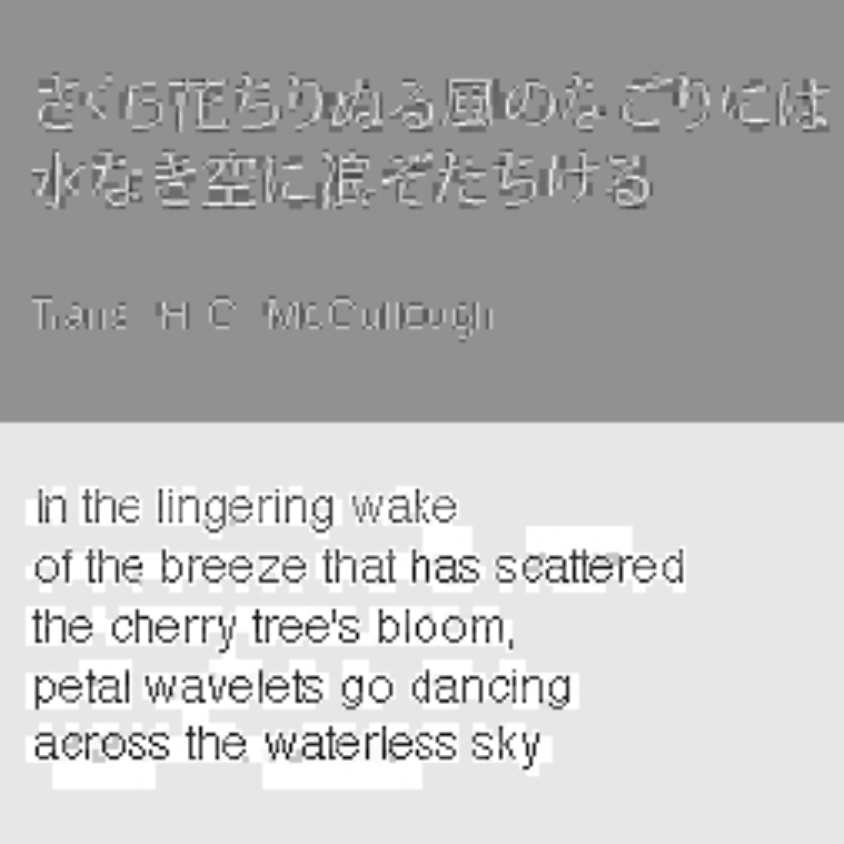}
\end{minipage}
} \subfigure[Output of our greedy algorithm under $\ell_1$]{\label{fig:L1poem}
\begin{minipage}[t]{2.83in}
\centering \includegraphics[width=2.83in]{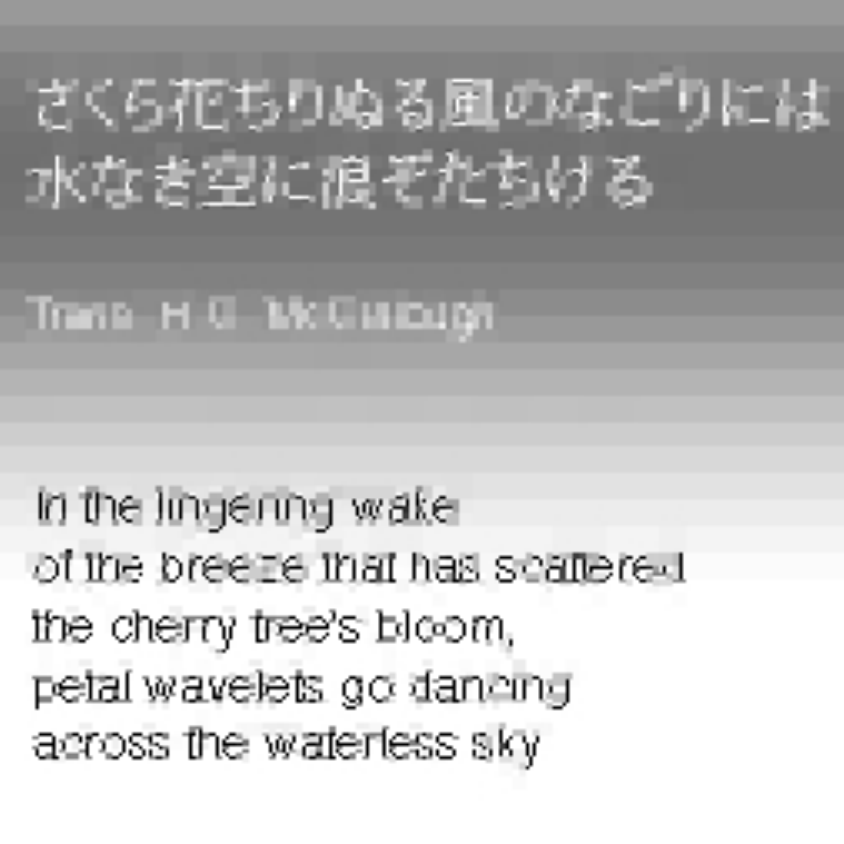}
\end{minipage}} 
\caption[Example image representation with the Haar wavelet using the
optimal $\ell_2$ strategy and our greedy approximation algorithm under
the $\ell_\infty$ and $\ell_1$ error measures] {Representing an image
  with embedded text using the optimal strategy that minimizes the
  $\ell_2$ error, and our greedy approximation algorithm under the
  $\ell_\infty$ and $\ell_1$ error measures.  The Haar wavelet is used
  in all three representations, and the number of retained
  coefficients is $B = 3840$.}
\label{fig:textonlightgradient}
\end{figure}

The first example illustrates a use of the $\ell_\infty$ measure for
sparse representation using wavelets. Minimizing the maximum error at
any point in the reconstructed image implies we should retain the
wavelet coefficients that correspond to sharp changes in intensity;
i.e., the coefficients that correspond to the ``details'' in the
image.  The image we used, shown in Figure~\ref{fig:originalpoem}, is
composed of a gradient background and both Japanese and English
texts\footnote{The Japanese text is poem number 89 of the
  \emph{Kokinshu} anthology~\cite{kokinshu}. The translation is by
  Helen Craig McCullough.}. The number of non-zero wavelet
coefficients in the original image is $65524$.  We set $B = 3840$ and
ran Algorithm~\ref{alg:daubgreedy} with $p=1,2$ and $\infty$ under the
Haar wavelet (with $\taps=1$). When $p=2$, the algorithm outputs the
optimal $B$-term representation that minimizes the $\ell_2$ error
measure. That is, the algorithm simply retains the largest $B$ wavelet
coefficients (since $\q=2$ and $\norm{\psi_i}_{\q} = 1$ for all
$i$). When $p=1$, or $p=\infty$, the algorithm outputs a $O(\log
n)$-approximate $B$-term representation as will be explained in
Section~\ref{sec:greedy}.  The results are shown in
Figure~\ref{fig:textonlightgradient}.  Notice that the $\ell_\infty$
representation essentially ignores the gradient in the background, and
it retains the wavelet coefficients that correspond to the text in the
image.  The $\ell_1$ representation also does better than the $\ell_2$
representation in terms of rendering the Japanese text; however, the
English translation in the former is not as clear.  The attribution in
the $\ell_2$ representation, on the other hand, is completely lost.
Although the differences between the three representations are not
stark, this example shows that under such high compression ratios
using the $\ell_\infty$ norm is more suitable for capturing signal
details than other norms.

\begin{figure}
\centering
\subfigure[The original image]{\label{fig:momoriginal}
\begin{minipage}[t]{2.83in}
\centering \includegraphics[width=2.83in]{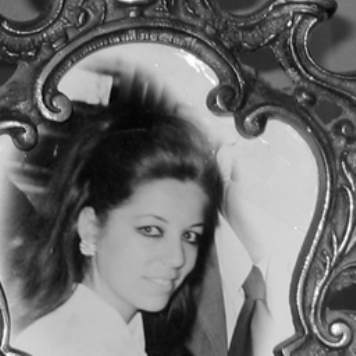}
\end{minipage}
} \subfigure[Output of the optimal $\ell_2$ algorithm 
(which retains the largest $B$ wavelet coefficients)]{\label{fig:L2mom}
\begin{minipage}[t]{2.83in}
\centering \includegraphics[width=2.83in]{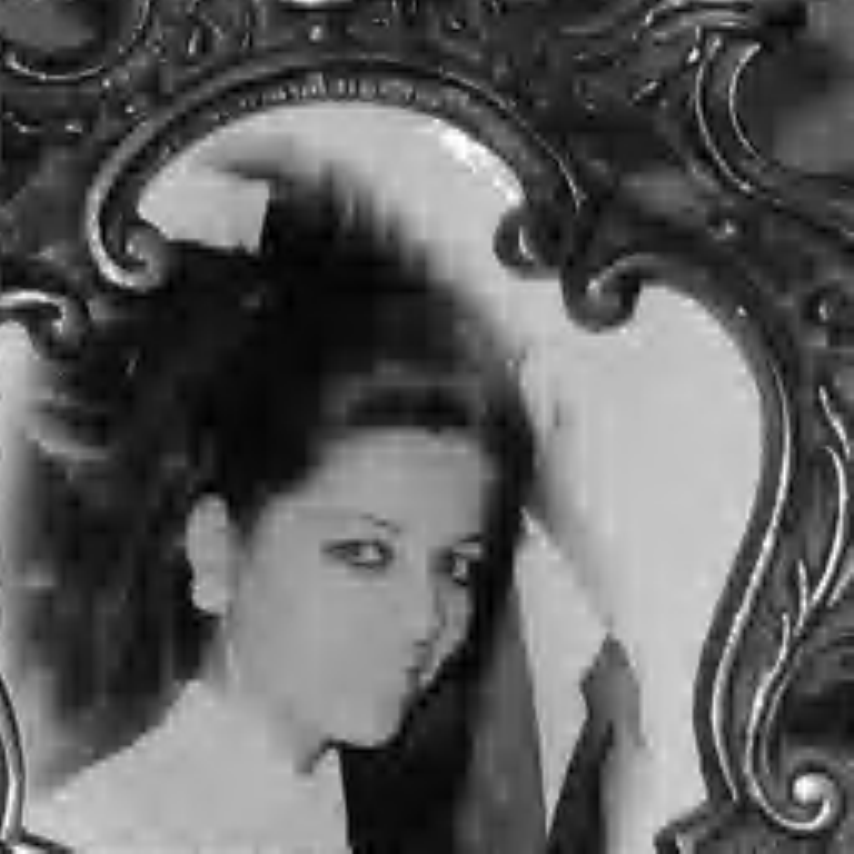}
\end{minipage}
} \\ \subfigure[Output of our greedy algorithm under $\ell_\infty$]{\label{fig:LInfmom}
\begin{minipage}[t]{2.83in}
\centering \includegraphics[width=2.83in]{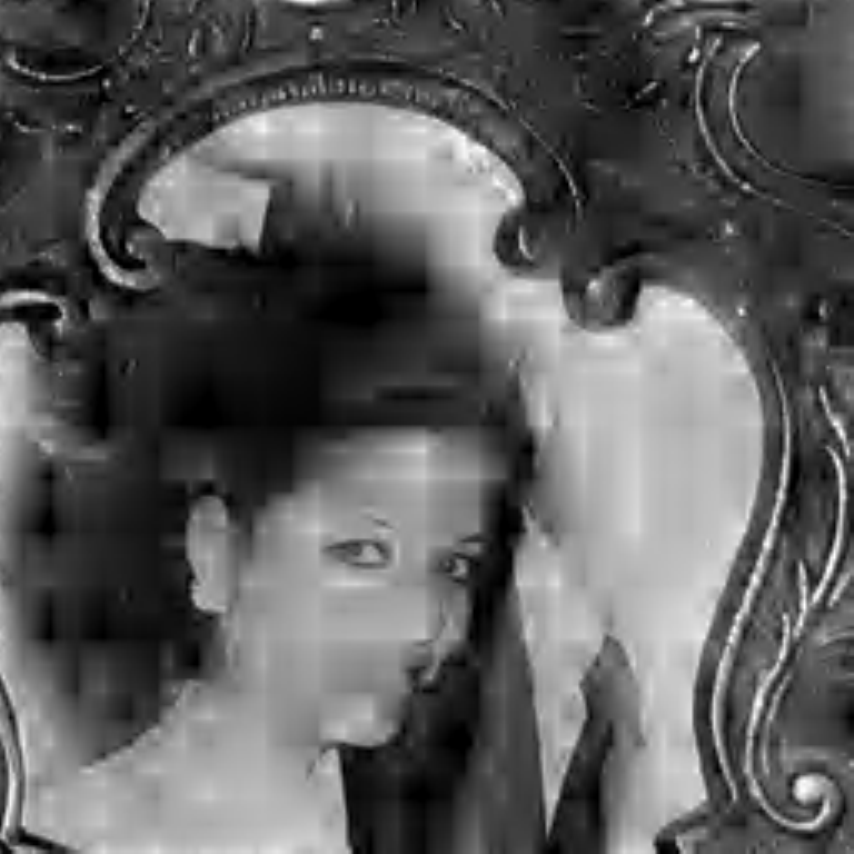}
\end{minipage}
} \subfigure[Output of our greedy algorithm under $\ell_1$]{\label{fig:L1mom}
\begin{minipage}[t]{2.83in}
\centering \includegraphics[width=2.83in]{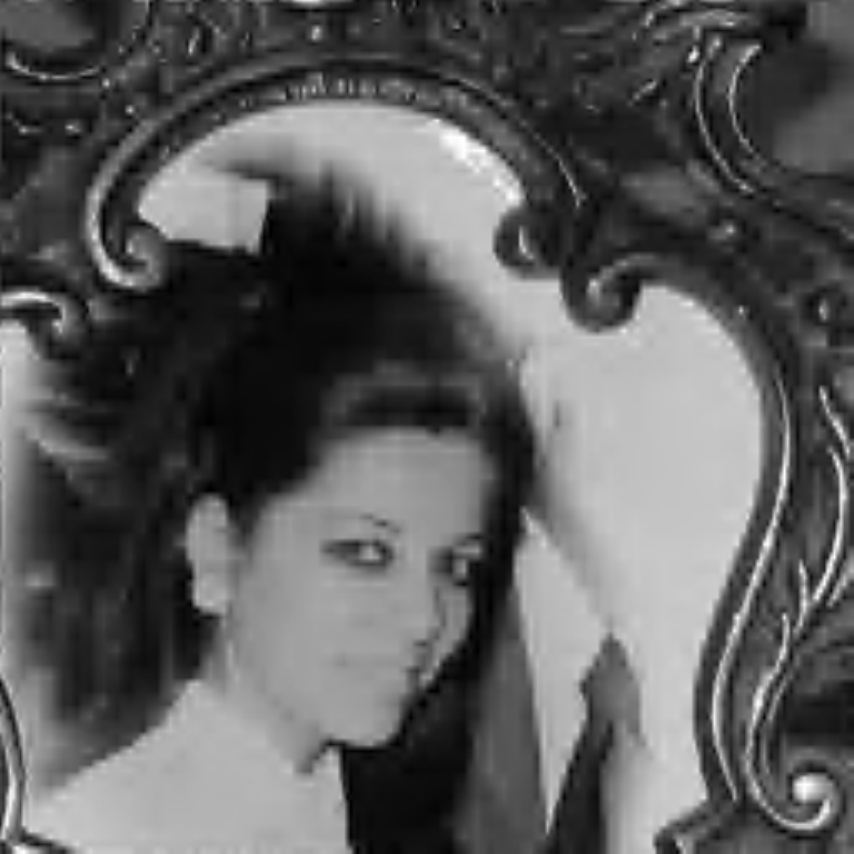}
\end{minipage}} 
\caption[Example image representation with the $D_2$ wavelet using the
optimal $\ell_2$ strategy and our greedy approximation algorithm under
the $\ell_\infty$ and $\ell_1$ error measures] {Representing an image
  using the optimal strategy that minimizes the $\ell_2$ error, and
  our greedy approximation algorithm under the $\ell_\infty$ and
  $\ell_1$ error measures.  The Daubechies $D_2$ wavelet is used in
  all three representations, and the number of retained coefficients
  is $B = 4096$.}
\label{fig:mom}
\end{figure}
\begin{figure}
\centering
\subfigure[The original image]{\label{fig:original_lamp}
\begin{minipage}[t]{2.83in}
\centering \includegraphics[width=2.83in]{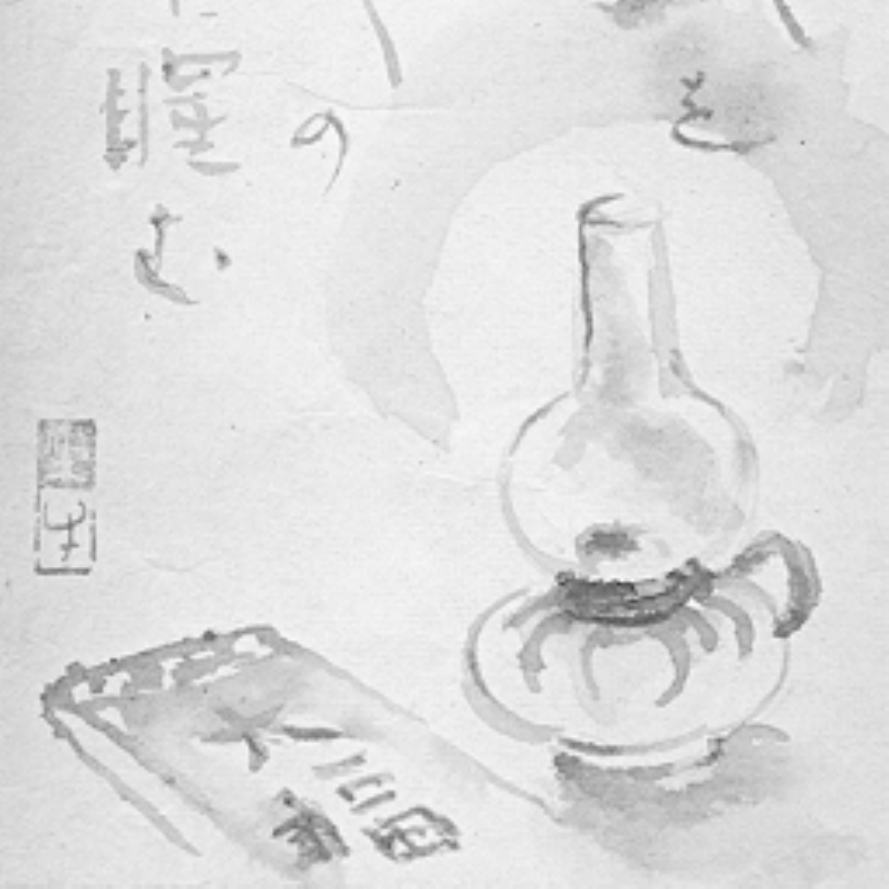}
\end{minipage}
} \subfigure[Output of the optimal $\ell_2$ algorithm 
(which retains the largest $B$ wavelet coefficients)]{\label{fig:L2lamp}
\begin{minipage}[t]{2.83in}
\centering \includegraphics[width=2.83in]{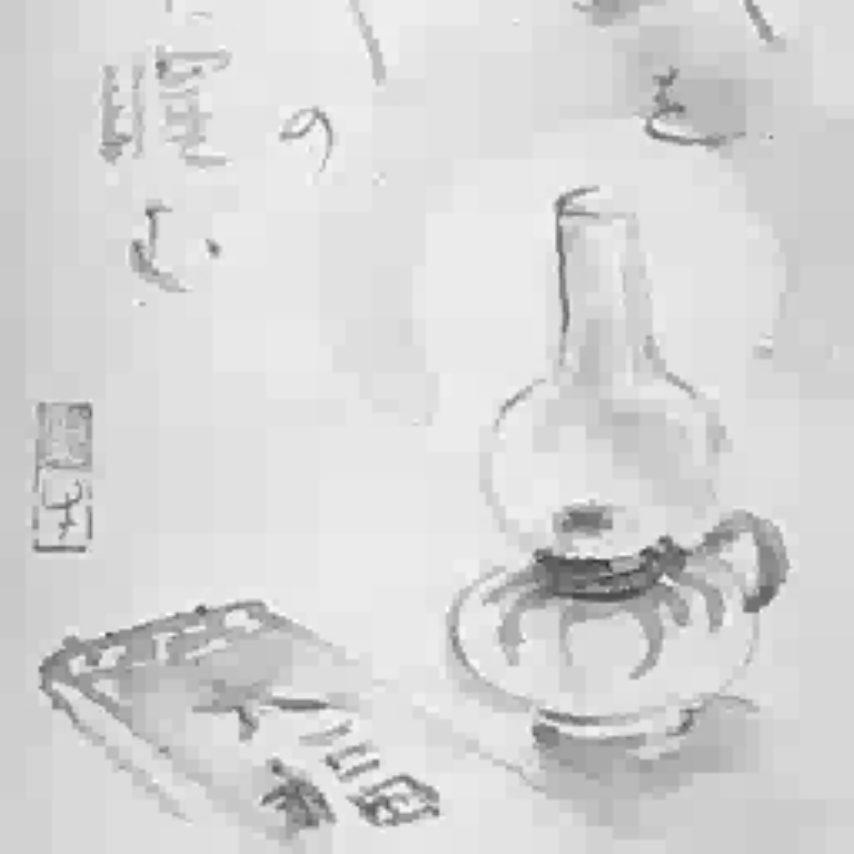}
\end{minipage}
} \\ \subfigure[Output of the best rank-$12$ approximation]{\label{fig:SVDlamp}
\begin{minipage}[t]{2.83in}
\centering \includegraphics[width=2.83in]{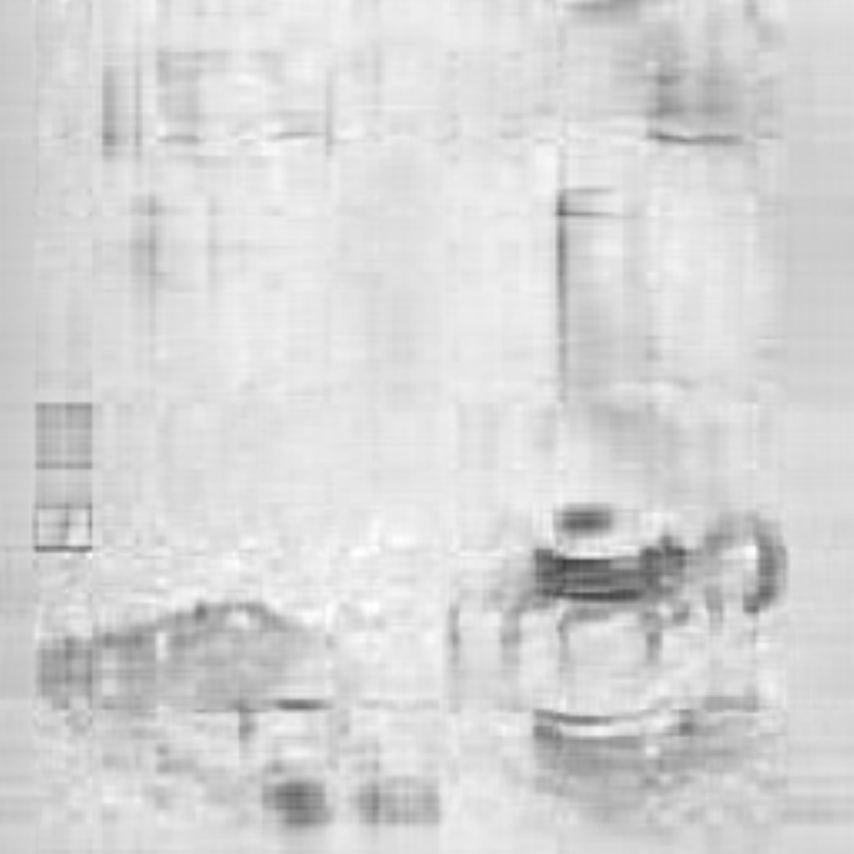}
\end{minipage}
} \subfigure[Output of our greedy algorithm under $\ell_1$]{\label{fig:L1lamp}
\begin{minipage}[t]{2.83in}
\centering \includegraphics[width=2.83in]{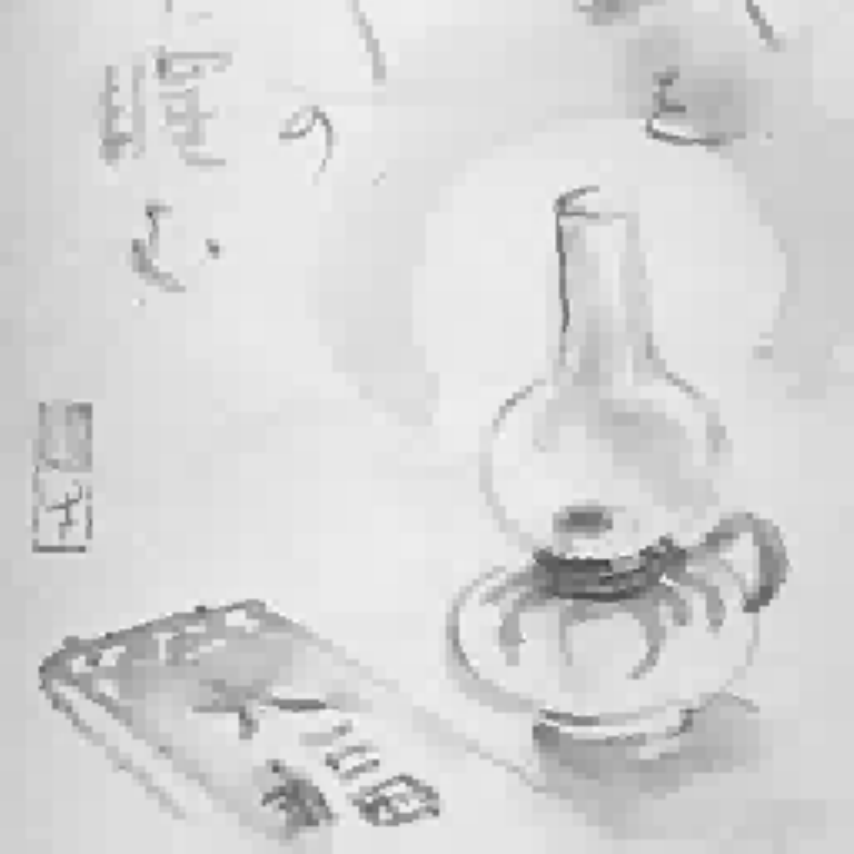}
\end{minipage}} 
\caption[Example image representation with the Haar wavelet using the
optimal $\ell_2$ strategy and our greedy approximation algorithm under
the $\ell_1$ error measure versus its best rank-$k$ approximation]
{Representing an image using the optimal strategy that minimizes the
  $\ell_2$ error and using our greedy approximation algorithm under
  the $\ell_1$ error measure versus its best rank-$k$
  approximation. Here $k=12$, and the number of values stored in all
  three representations is $6144$.  The Haar wavelet is used in the
  two nonlinear representations (the number of retained wavelet
  coefficients is $B = 3072$).}
\label{fig:lamp}
\end{figure}

The second example illustrates a use of the $\ell_1$ error measure.
Since the $\ell_1$ norm is robust in the sense that it is indifferent
to outliers, the allocation of wavelet coefficients when minimizing
the $\ell_1$ norm will be less sensitive to large changes in intensity
than the allocation under the $\ell_2$ norm.  In other words, it
implies that under the $\ell_1$ norm the wavelet coefficients will be
allocated more evenly across the image. The image we used, shown in
Figure~\ref{fig:momoriginal}, is a framed black and white matte
photograph. The number of non-zero wavelet coefficients in the
original image is $65536$. We set $B = 4096$ and ran
Algorithm~\ref{alg:daubgreedy} with $p=1,2$ and $\infty$ under the
Daubechies $D_2$ wavelet.  The results are shown in
Figure~\ref{fig:mom}.  Notice that the face of the subject is rendered
in the $\ell_1$ representation more ``smoothly'' than in the $\ell_2$
representation.  Further, the subject's mouth is not portrayed
completely in the $\ell_2$ representation. As explained earlier, these
differences between the two representations are due to the fact that
the $\ell_1$ norm is not as affected as the $\ell_2$ norm by other
conspicuous details in the image; e.g., the frame.  The $\ell_\infty$
representation, on the other hand, focuses on the details of the image
displaying parts of the frame and the eyes well, but misses the rest
of the subject entirely. This example foregrounds some advantages of
the $\ell_1$ norm over the customary $\ell_2$ norm for compressing
images.

The last example highlights the advantage of representing an image
sparsely using a nonlinear wavelet approximation versus using a
rank-$k$ approximation of the image.  Recall that if $X$ is our image
then the best rank-$k$ approximation is given by $U_k \Sigma_k V_k^T$
where $X = U\Sigma V^T$ is the SVD decomposition of $X$, and $U_k$ is
comprised of the $k$ singular vectors corresponding to the largest $k$
singular values of $X$ (see, e.g.,~\cite{GVL89}).  The original image
is shown in Figure~\ref{fig:original_lamp}\footnote{The image is taken
  from a water painting by Shozo Matsuhashi.  It is untitled.} and the
number of non-zero coefficients in its Haar wavelet expansion is
$65536$.  Figure~\ref{fig:SVDlamp} shows the best rank-$12$
approximation of the image; i.e., it displays $X_{12} =
U_{12}\Sigma_{12}V^T_{12}$.  This representation stores $6144$ values
corresponding to the number of elements in $U_{12}\Sigma_{12}$ plus
$V_{12}$.  We set $B = 3072$ and ran Algorithm~\ref{alg:daubgreedy}
with $p=1,2$ under the Haar wavelet (Figures~\ref{fig:L1lamp}
and~\ref{fig:L2lamp}). (The $B$-term representation problem implicitly
requires storing $2B$ numbers: the $B$ \emph{values} of the solution
components that we compute, and the $B$ indices of these components.)
It is clear that the nonlinear approximations offer perceptually
better representations that the approximation offered by the SVD.
Also, as in the previous example, the $\ell_1$ representation is again
``smoother'' than the $\ell_2$ with less visible artifacts.

\section{A Streaming $(1+\epsilon)$ Approximation for Haar Wavelets}
\label{apxschemes}
In this section we will provide a FPTAS for the Haar system.  The
algorithm will be bottom up, which is convenient from a streaming
point of view.  Observe that in case of general $\ell_p$ norm error,
we cannot disprove that the optimum solution cannot have an irrational
value, which is detrimental from a computational point of view.  In a
sense we will seek to narrow down our search space, but we will need
to preserve near optimality.  We will show that {\em there exists}
sets $R_i$ such that if the solution coefficient $z_i$ was drawn from
$R_i$, then {\em there exists} one solution which is close to the
optimum unrestricted solution (where we search over all reals).  In a
sense the sets $R_i$ ``rescue'' us from the search. Alternately we can
view those sets as a ``rounding'' of the optimal solution.  Obviously
such sets exist if we did not care about the error, e.g. take the all
zero solution. We would expect a dependence between the sets $R_i$ and
the error bound we seek.  We will use a type of ``dual'' wavelet
bases; i.e., where we use one basis to construct the coefficients and
another to reconstruct the function. Our bases will differ by scaling
factors.  We will solve the problem in the scaled bases and translate
the solution to the original basis.  This overall approach is similar
to that in \cite{GH05}, however, it is different in several details
critical to the proofs of running time, space complexity and
approximation guarantee.

\begin{Definition}\label{def:psi-ab}
Define $\psia_{j,s}=2^{-j/2}\psi_{j,s}$ and
$\psib_{j,s}=2^{j/2}\psi_{j,s}$. 
Likewise define $\phia_{j,s} = 2^{-j/2}\phi_{j,s}$.
\end{Definition}

\begin{proposition}
The Cascade algorithm used with $\frac1{\sqrt{2}}h[]$ computes 
$\langle f, \psia_i \rangle$ and $\langle f,\phia_i\rangle$.
\end{proposition}

\noindent We now use the change of basis. The next proposition is
clear from the definition of $\{\psi^b_i\}$.

\begin{proposition}
The problem of finding a representation $\hat{f}$ with $\{z_i\}$ and
basis $\{\psi_i\}$ is equivalent to finding the same representation
$\hat{f}$ using the coefficients $\{y_i\}$ and the basis $\{\psib_i\}$.  
The correspondence is $y_i = y_{j,s} = 2^{-j/2}z_{j,s}$.
\hide{and there are no more than $B$ non-zero $y_i$'s if and only if
there are no more than $B$ non-zero $z_i$.}
\end{proposition}

\begin{lemma}
\label{changebase1}
Let $\{ y^*_i\}$ be the optimal solution using the basis set
$\{\psib_i\}$ for the reconstruction, i.e., $\hat{f} = \sum_i
y^*_i\psib_i$ and $\| f - \hat{f}\|_p = \E$. Let $\{y^\rho_i\}$ be the
set where each $y^*_i$ is rounded to the nearest multiple of
$\rho$. If $f^\rho = \sum_i y^\rho_i\psib_i$ then $\|f -
f^\rho\|_p \leq \E + O(qn^{1/p}\rho\log n)$.
\end{lemma}
\begin{proof}
Let $\rho_i = y^*_i - y^\rho_i$.  By the triangle inequality, 
\[ \|f - f^\rho\|_p \leq \E + \norm{\sum\nolimits_i \rho_i\psib_i}_p \enspace .\]
Proposition~\ref{prop:qlogn-basis} and the fact that $\abs{\rho_i} \le \rho$
imply $\abs{\sum_k\rho_i\psib_i(k)} \le c\rho q\log n \max_i\abs{\psib_i(k)}$ 
for a small constant $c$.  This bound gives 
$\|f - f^\rho\|_p \leq \E + O(qn^{1/p}\rho\log  n \max_i \|\psib_i\|_\infty)$.
Now $\psib_i = \psib_{j,s} = 2^{j/2}\psi_{j,s}$, and from the proof of
Lemma~\ref{second} we know that for large $j$, $\|\psi_{j,s}\|_\infty$ 
is at most $2^{-j/2}$ times a constant. 
For smaller $j$, $\|\psib_{j,s}\|_\infty$ is a constant.
\end{proof}

We will provide a dynamic programming formulation using the new
basis. But we still need to show two results; the first concerning the
$y^*_i$'s and the second concerning the $a_j[]$'s. The next lemma is
very similar to Lemma~\ref{lb} and follows from the fact that
$\|\psia_{j,s}\|_1 = 2^{-j/2}\|\psi_{j,s}\|_1 \le \sqrt{2q}$.
\begin{lemma}
\label{psilemma}
$ - C_0\sqrt{q}\E \leq \langle f, \psia_i \rangle - y^*_i \leq C_0\sqrt{q}\E$ 
for some constant $C_0$.
\end{lemma}
\hide{ %%% Proof is very similar to Lemma \ref{lb}.
\begin{proof}
We can follow the proof of Lemma~\ref{lb} and use the fact that if
$i=(j,s)$ we have $\langle \psia_{i}, \psi_{k} \rangle =
2^{-j/2}\delta_{ik}$. The only other thing we need to show is that
$\|\psia_i\|_1$ is a constant. This
follows from the proof of Lemma~\ref{second}, where we show that
$\|\psi_i\|_1$ is $O(2^{j/2})$ if $i$ is of scale $j$. Since
$\psia_i=2^{-j/2}\psi_i$ the lemma follows.
\end{proof}
}
Now suppose we know the optimal solution $\hat{f}$, and suppose we are
computing the coefficients $a_j[]$ and $d_j[]$ for both $f$ and
$\hat{f}$ at each step $j$ of the Cascade algorithm.  We wish to know
by how much their coefficients differ since bounding this gap would
shed more light on the solution $\hat{f}$.

\begin{proposition}
  Let $a_j[s](F)$ be $a_j[s]$ computed from $a_0[s]=F(s)$ then
  $a_j[s](f)-a_j[s](\hat{f})=a_j[s](f-\hat{f})$.
\end{proposition}

\begin{lemma}
\label{philemma}
  If $\|f -\hat{f}\|_p \leq\E$ then $|a_j[s](f-\hat{f})|\leq C_1\sqrt{q}\E$
  for some constant $C_1$. (We are using $\frac{1}{\sqrt2} h[]$.)  
\end{lemma}
\begin{proof}
The proof is similar to that of Lemma~\ref{lb}.
Let $F=f-\hat{f}$. We know $-\E \leq F(i) \leq
\E$. Multiplying by $|\phia_{j,s}(i)|$ and summing over all $i$ we get
$ -\E \|\phia_{j,s}\|_1 \leq \langle F, \phia_{j,s} \rangle =
a_j[s](F) \leq \E \|\phia_{j,s}\|_1$.  By definition,
$\phia_{j,s}=2^{-j/2}\phi_{j,s}$. Further, $\|\phi_{j,s}\|_2=1$ and
has at most $(2q)2^j$ non-zero values. 
Hence, $\|\phia_{j,s}\|_1 \leq \sqrt{2q}$.  The lemma follows.
\end{proof}
At this point we have all the pieces. Summarizing:
\begin{lemma}\label{lemma:summary}
Let $\{z_i\}$ be a solution with $B$ non-zero coefficients and with
representation $\hat{f}=\sum_i z_i \psi_i$.  
If $\|f-\hat{f}\|_p \leq \E$, then there is a solution $\{y_i\}$ with
$B$ non-zero coefficients and representation $f'=\sum_i y_i \psib_i$
such that for all $i$ we have,
\begin{enumerate}
\item[(i)] $y_i$ is a multiple of $\rho$;
\item[(ii)] $|y_i - \langle f,\psia_{i} \rangle | \leq C_0\sqrt{q}\E + \rho$; and,
\item[(iii)] $| \langle f,\phia_i \rangle - \langle f',\phia_i\rangle| \leq C_1\sqrt{q}\E +O(q\rho \log n)$,
\end{enumerate}
and $\|f -f'\|_p \leq \E + O(qn^{1/p}\rho \log n)$.
\end{lemma}
\begin{proof}
Rewrite $\hat{f}=\sum_i z_i \psi_i = \sum_i z_i^*\psib_i$ where 
$z_i^* = z_{j,s}^* = 2^{-j/2} z_{j,s}$. Let $\{y_i\}$ be the
solution where each $y_i$ equals $z^*_i$ rounded to the nearest multiple of
$\rho$. Lemmas~\ref{psilemma} and~\ref{philemma} bound the $z_i^*$'s thus
providing properties (ii) and (iii). Finally, Lemma~\ref{changebase1}
gives the approximation guarantee of $\{y_i\}$.
\end{proof}

The above lemma ensures the existence of a solution $\{y_i\}$ that is
$O(qn^{1/p}\rho \log n)$ away from the optimal solution and that
possesses some useful properties which we shall exploit for designing
our algorithms.  Each coefficient $y_i$ in this solution is a multiple
of a parameter $\rho$ that we are free to choose, and it is a constant
multiple of $\E$ away from the $i^\text{th}$ wavelet coefficient of
$f$.  Further, without knowing the values of those coefficients
$y_{j,s}$ contributing to the reconstruction of a certain point
$f'(i)$, we are guaranteed that during the incremental reconstruction
of $f'(i)$ using the cascade algorithm, every $a_j[s](f')$ in the
support of $f'(i)$ is a constant multiple of $\E$ away from $a_j[s](f)
= \langle f, \phia_{j,s}\rangle$.  This last property allows us to
design our algorithms in a bottom-up fashion making them suitable for
data streams.  Finally, since we may choose $\rho$, setting it
appropriately results in true factor approximation algorithms. Details
of our algorithms follow.

\subsection{The Algorithm: A Simple Version}\label{sec:HaarAlgo}
We will assume here that we know the optimal error $\E$.  This
assumption can be circumvented by running $O(\log n)$ instances of the
algorithm presented below `in parallel', each with a different guess
of the error.  This will increase the time and space requirements of
the algorithm by a $O(\log n)$ factor, which is accounted for in
Theorem~\ref{mainthm} (and also in Theorem~\ref{mainthm2}). We detail
the guessing procedure in Section~\ref{sec:guesses}.  Our algorithm
will be given $\E$ and the desired approximation parameter $\epsilon$
as inputs (see Fig.~\ref{fig:apx}). 
\medskip

The Haar wavelet basis naturally form a complete binary tree, termed
the \emph{coefficient tree}, since their support sets are nested and
are of size powers of $2$ (with one additional node as a parent of the
tree). The data elements correspond to the leaves, and the
coefficients correspond to the non-leaf nodes of the tree. Assigning a
value $y$ to the coefficient corresponds to assigning $+y$ to all the
leaves that are {\em left descendants} (descendants of the left child)
and $-y$ to all right descendants (recall the definition of
$\{\psib_i\}$).  The leaves that are descendants of a node in the
coefficient tree are termed the {\em support} of the coefficient.

\begin{Definition}
Let $E[i,v,b]$ be the minimum possible contribution to the overall
error from all descendants of node $i$ using exactly $b$ coefficients,
under the assumption that ancestor coefficients of $i$ will add up to
the value $v$ at $i$ (taking account of the signs) in the final
solution.
\end{Definition}

The value $v$ will be set later for a subtree as more data
arrive. Note that the definition is bottom up and after we compute the
table, we do not need to remember the data items in the subtree. As
the reader would have guessed, this second property will be
significant for streaming.

The overall answer is $\min_b E[root,0,b]$---by the time we are at the
root, we have looked at all the data and no ancestors exist to set a
non-zero $v$. A natural dynamic program arises whose idea is as
follows: Let $i_L$ and $i_R$ be node $i$'s left and right children
respectively.  In order to compute $E[i,v,b]$, we guess the
coefficient of node $i$ and minimize over the error produced by $i_L$
and $i_R$ that results from our choice.  Specifically, the computation
is:

\begin{enumerate}
\item A non-root node computes $E[i,v,b]$ as follows:
\vspace{-0.05in}
\[ \min \left \{ \begin{array}{l}
\min_{r,b'} E[i_L,v+r,b'] + E[i_R,v-r,b-b'-1] \\
\min_{b'} E[i_L,v,b'] + E[i_R,v,b-b']
\end{array} \right.
\]
where the upper term computes the error if the $i^{th}$ coefficient is
chosen and it's value is $r\in R_i$ where $R_i$ is the set of
multiples of $\rho$ between $\langle f, \psia_i\rangle -
C_0\sqrt{q}\E$ and $\langle f, \psia_i\rangle + C_0\sqrt{q}\E$; and
the lower term computes the error if the $i^{th}$ coefficient is not
chosen.

\item  Then the root node computes: 
\[ \min \left \{
\begin{array}{ll}
\min_{r,b'} E[i_C,r,b'-1] & \mbox{root coefficient is $r$}\\
\min_{b'} E[i_C,0,b'] & \mbox{root not chosen}
\end{array} \right.
\]
where $i_C$ is the root's only child.
\end{enumerate}

The streaming algorithm will
borrow from the paradigm of reduce-merge. The high level idea
will be to construct and maintain a small table of possibilities
for each resolution of the data. On seeing each item $f(i)$, we
will first find out the best choices of the wavelets of length one
(over all future inputs) and then, if appropriate,
construct/update a table for wavelets of length $2,4,\ldots$ etc.

The idea of subdividing the data, computing some information and
merging results from adjacent divisions were used in \cite{GMMO00}
for stream clustering. The stream computation of wavelets in
\cite{GKMS01} can be viewed as a similar idea---where the
divisions corresponds to the support of the wavelet basis vectors.

\begin{figure}
\centering
\subfigure[The arrival of the first $3$ elements.]{\label{fig:salg123}
\begin{minipage}[t]{1.2in}
\centering \includegraphics[width=1in]{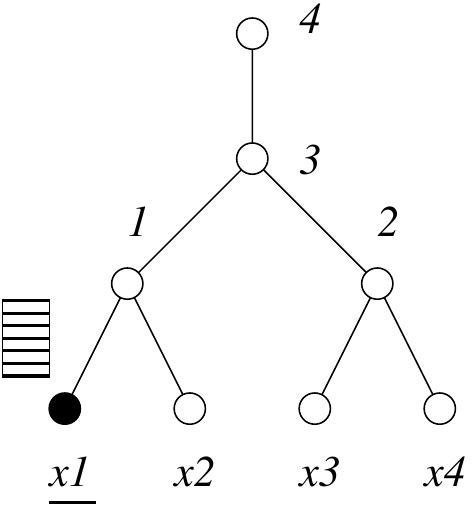}
\end{minipage}
\begin{minipage}[t]{1.2in}
\centering \includegraphics[width=1in]{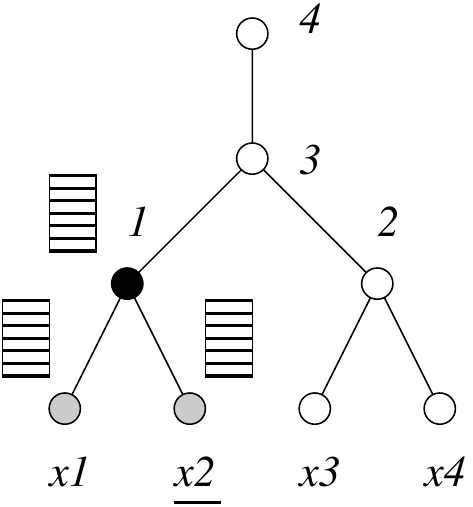}
\end{minipage}
}  \subfigure[The arrival of $x_4$]{\label{fig:salg456}
\begin{minipage}[t]{1.2in}
\centering \includegraphics[width=1in]{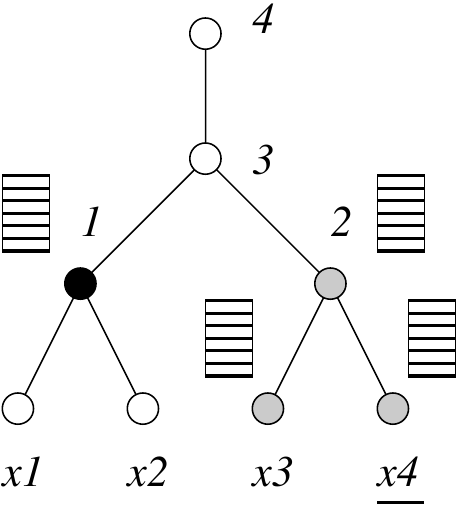}
\end{minipage}
\begin{minipage}[t]{1.2in}
\centering \includegraphics[width=1in]{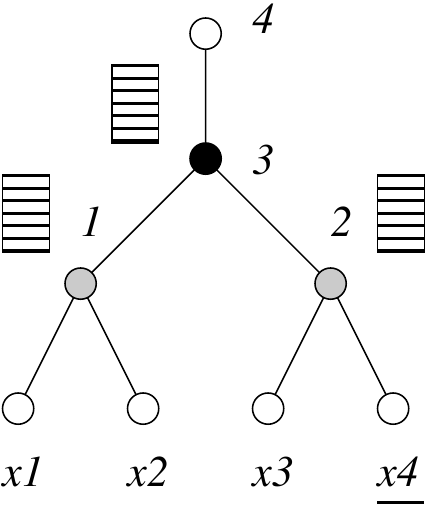}
\end{minipage}}
\caption{Upon seeing $x_2$ node $1$ computes
$\mbox{$E[1,\cdot,\cdot]$}$ and the two error arrays associated with
$x_1$ and $x_2$ are discarded.  Element $x_4$ triggers the computation
of $\mbox{$E[2, \cdot, \cdot ]$}$ and the two error arrays associated
with $x_3$ and $x_4$ are discarded. Subsequently, $\mbox{$E[3,\cdot,
\cdot ]$}$ is computed from $\mbox{$E[1,\cdot,\cdot]$}$ and
$\mbox{$E[2,\cdot,\cdot ]$}$ and both the latter arrays are
discarded. If $x_4$ is the last element on the stream, the root's
error array, $\mbox{$E[3,\cdot,\cdot ]$}$, is computed from
$\mbox{$E[2,\cdot,\cdot]$}$.}
\end{figure}

Our streaming algorithm will compute the error arrays
$E[i,\cdot,\cdot]$ associated with the internal nodes of the coefficient
tree in a post-order fashion. Recall that the wavelet basis
vectors, which are described in Section~\ref{sec:prelim}, form a
complete binary tree. For example, the scaled basis vectors for nodes $4,
3, 1$ and $2$ in the tree of Fig.~\ref{fig:salg123} are
$[1,1,1,1]$, $[1,1,-1,-1]$, $[1,-1,0,0]$ and $[0,0,1,-1]$
respectively. The data elements correspond to the leaves of the
tree and the coefficients of the synopsis correspond to its
internal nodes. 
\eat{
Hence, assigning the value $c$ to node $2$ (equivalently, setting
$z_2=c$) for example corresponds to adding $c$ to $\wai(Z)_1$ and
$\wai(Z)_2$, and adding $-c$ to $\wai(Z)_3$ and $\wai(Z)_4$.
}

We need not store the error array for every internal node since, in
order to compute $E[i,v,b]$ our algorithm only requires that
$E[i_L,\cdot,\cdot ]$ and $E[i_R,\cdot,\cdot ]$ be known.  Therefore,
it is natural to perform the computation of the error arrays in a
post-order fashion. An example best illustrates the procedure. Suppose
$f = \langle x_1,x_2,x_3,x_4\rangle$. In Fig.~\ref{fig:salg123} when
element $x_1$ arrives, the algorithm computes the error array
associated with $x_1$, call it $E_{x_1}$.  When element $x_2$ arrives
$E_{x_2}$ is computed.  The array $E[1,\cdot,\cdot ]$ is then computed
and $E_{x_1}$ and $E_{x_2}$ are discarded. Array $E_{x_3}$ is computed
when $x_3$ arrives.  Finally the arrival of $x_4$ triggers the
computations of the rest of the arrays as in Fig.~\ref{fig:salg456}.
Note that at any point in time, there is only one error array stored
at each \emph{level} of the tree.  In fact, the computation of the
error arrays resembles a binary counter.  We start with an empty queue
$Q$ of error arrays. When $x_1$ arrives, $E_{q_0}$ is added to $Q$ and
the error associated with $x_1$ is stored in it.  When $x_2$ arrives,
a temporary node is created to store the error array associated with
$x_2$.  It is immediately used to compute an error array that is added
to $Q$ as $E_{q_1}$. Node $E_{q_0}$ is emptied, and it is filled again
upon the arrival of $x_3$. When $x_4$ arrives: (1) a temporary
$E_{t_1}$ is created to store the error associated with $x_4$; (2)
$E_{t_1}$ and $E_{q_0}$ are used to create $E_{t_2}$; $E_{t_1}$ is
discarded and $E_{q_0}$ is emptied; (3) $E_{t_2}$ and $E_{q_1}$ are
used to create $E_{q_2}$ which in turn is added to the queue;
$E_{t_2}$ is discarded and $E_{q_1}$ is emptied.  
The algorithm  for $\ell_\infty$ is shown in Fig.~\ref{fig:apx}.

%\begin{figure*}[htb]
\clearpage
\begin{figure}
\framebox[6.7in]{\parbox{6.5in}{
\begin{algorithm}{HaarPTAS}[B,\E,\epsilon]{\label{alg:apx}}
Let $\rho = \epsilon\E/(c q \log n)$ for some suitably
large constant $c$.  Note that $q=1$ in the Haar case.\\
Initialize a queue $Q$ with one node $q_0$ \qcomment{Each $q_i$
contains an array $E_{q_i}$ of size at most
$R\min\{B, 2^i\}$ and a flag {\tt isEmpty}}\\
{\bf repeat} Until there are no elements in the stream\\
Get the next element from the stream, call it $e$\\
\qif $q_0$ is empty \\
\qthen Set $q_0.a = e$. For all values $r$ s.t.~$|r -e| \leq c_1 \E$
  where $c_1$ is a large enough constant and $r$ is a multiple of
  $\rho$, initialize the table $E_{q_0}[r, 0] =
  |r-e|$\label{step:baseE} \\
\qelse Create $t_1$ and Initialize $E_{t_1}[r, 0] =|r-e|$ \emph{as in
Step \ref{step:baseE}}.\\
\qfor $i=1$ until the $1^\text{st}$ empty $q_i$ or end of $Q$ \\
\qdo Create a temporary node $t_2$.\\
Compute $t_2.a = \langle f,\phia_i\rangle$ and the wavelet coefficient
$t_2.o=\langle f, \psia_i\rangle$. This involves using the $a$ values
of $t_{1}$ and $q_{i-1}$ ($t_2$'s two children in the coefficient
tree) and taking their average to compute $t_2.u$ and their difference
divided by $2$ to compute $t_2.o$. (Recall that we are using
$\frac{1}{\sqrt{2}}h[]$).\\
For all values $r$ that are multiples of $\rho$ with $|r -t_2.a| \leq
  c_1(\E + \rho\log n)$, compute the table $E_{t_2}[r, b]$ for all $0\leq b \leq
  B$. This uses the tables of the two children $t_{1}$ and
  $q_{i-1}$. The size of the table is $O(\epsilon^{-1}Bn^{1/p}\log
  n)$. (Note that the value of a chosen coefficient at node $t_2$ is at
  most a constant multiple of $\E$ away from $t_2.o$. Keeping track of
  the chosen coefficients (the answer) costs $O(B)$ factor space
  more.)\label{step:generalE}\\
Set $t_1 \leftarrow t_2$ and Discard $t_2$\\
Set $q_i.\mathtt{isEmtpy} = \mbox{true}$
\qrof \\
\qif we reached the end of $Q$ \\
\qthen Create the node $q_i$ \qfi \\
Compute $E_{q_i}[r, b\in B]$ from $t_{1}$ and $q_{i-1}$ \emph{as in
Step \ref{step:generalE}}.\\
Set $q_i.\mathtt{isEmpty} = \mbox{false}$ and Discard $t_{1}$ \qfi
\end{algorithm}
}}
\caption{The Haar streaming FPTAS for $\ell_\infty$.}
\label{fig:apx}
\end{figure}
\clearpage

%If at any point of time the number of coefficients larger than $\E$
%exceeds $B$ then we know our guess of $\E$ is wrong and we abort that
%thread.
  \subsubsection{Guessing the Optimal Error}\label{sec:guesses}
We have so far assumed that we know the optimal error $\E$. As
mentioned at the beginning of Section~\ref{sec:HaarAlgo}, we will
avoid this assumption by running multiple instances of our algorithm
and supplying each instance a different guess $G_k$ of the error.  We
will also provide every instance $A_k$ of the algorithm with
$\epsilon' = \frac{\sqrt{1+4\epsilon}-1}{2}$ as the approximation
parameter.  The reason for this will be apparent shortly.  Our final
answer will be that of the instance with the minimum representation
error.

Theorem~\ref{mainthm} shows that the running time and space
requirements of our algorithm do not depend on the supplied error
parameter.  However, the algorithm's search ranges {\it do} depend on
the given error. Hence, as long as $G_k\ge\E$ the ranges searched by
the $k^\text{th}$ instance will include the ranges specified by
Lemma~\ref{lemma:summary}.  Lemma~\ref{lemma:summary} also tells us
that if we search these ranges in multiples of $\rho$, then we will
find a solution whose approximation guarantee is $\E+ c q
n^{1/p}\rho\log n$.  Our algorithm chooses $\rho$ so that its running
time does not depend on the supplied error parameter.  Hence, given
$G_k$ and $\epsilon'$, algorithm $A_k$ sets $\rho = \epsilon'G_k/(c q
n^{1/p}\log n)$.  Consequently, its approximation guarantee is $\E +
\epsilon' G_k$.

Now if guess $G_k$ is much larger than the optimal error $\E$, then
instance $A_k$ will not provide a good approximation of the optimal
representation.  However, if $G_k \le (1+\epsilon')\E$, then $A_k$'s
guarantee will be $\E+ \epsilon'(1+\epsilon')\E = (1+\epsilon)\E$
because of our choice of $\epsilon'$.  To summarize, in order to
obtain the desired $(1+\epsilon)$ approximation, we simply need to
ensure that one of our guesses (call it $G_{k^*}$) satisfies
\begin{equation*}\label{eq:guess}
\E \le\ G_{k^*} \le\ (1+\epsilon')\E
\end{equation*}
Setting $G_k = (1+\epsilon')^k$, the above bounds will be satisfied
when 
$k = k^* \in [\log_{1+\epsilon'}(\E),\ \log_{1+\epsilon'}(\E) +1]$.  

\paragraph*{Number of guesses}
Note that the optimal error $\E = 0$ if and only if $f$ has at 
most $B$ non-zero expansion coefficients $\langle f, \psi_i\rangle$. 
We can find these coefficients easily in a streaming fashion.

Since we assume that the entries in the given $f$ are polynomially
bounded, by the system of equations~\eqref{sys} we know that the
optimum error is at least as much as the $(B+1)^{\text{st}}$ largest
coefficient. Now any coefficient ($\langle f, \psia_k\rangle$) is the
sum of the left half minus the sum of the right half of the $f_i$'s
that are in the support of the basis and the total is divided by the
length of the support. Thus if the smallest non-zero number in the
input is $n^{-c}$ then the smallest non-zero wavelet coefficient is at
least $n^{-(c+1)}$. By the same logic the largest non-zero coefficient
is $n^c$.  Hence, it suffices to make $O(\log n)$ guesses.

\medskip
\subsection{Analysis of the Simple Algorithm}
\label{sec:algspacetime}
The size of the error table at node $i$, $E[i,\cdot,\cdot]$, is
$R_\phi \min\{B, 2^{t_i}\}$ where $R_\phi = 2C_1\E/\rho+\log n$ and $t_i$ is
the height of node $i$ in the Haar coefficient tree (the leaves have
height $0$). Note that $q=1$ in the Haar case.  Computing each entry
of $E[i,\cdot,\cdot]$ takes $O(R_\psi\min\{B, 2^{t_i}\})$ time where
$R_\psi = 2C_0\E/\rho+2$. Hence, letting $R = \max\{R_\phi, R_\psi\}$,
the total running time is $O(R^2B^2)$ for computing the root table
plus $O(\sum_{i=1}^n \left(R\min \{ 2^{t_i},B\}\right)^2)$ for
computing all the other error tables. Now,
\begin{eqnarray*}
\sum_{i=1}^n \left(R \min \{ 2^{t_i},B \}\right)^2
& = & R^2 \sum_{t=1}^{\log n} \frac{n}{2^t} \min \{ 2^{2t},B^2\} \\
& = & nR^2\left(\sum_{t=1}^{\log B}2^t + \sum_{t=\log B +1}^{\log n} \frac{B^2}{2^t}\right) \\
%&=& n|R|^2\left((2B-2) + \sum_{u=1}^{\log (n/B)}\frac{B}{2^{u}}\right)\\
& = & O(R^2nB) \enspace ,
\end{eqnarray*}
where the first equality follows from the fact that the number of
nodes at level $t$ is $\frac{n}{2^t}$. For $\ell_\infty$, when
computing $E[i,v,b]$ we do not need to range over all values of
$B$. For a specific $r\in R_i$, we can find the value of $b'$ that
minimizes $\max\{E[i_L,v+r,b'], E[i_R,v-r,b-b'-1]\}$ using binary
search. The running time thus becomes,
\[
\sum_{t} R^2 \frac{n}{2^t} \min \{t2^{t},B \log B \} = O(nR^2\log^2 B) \enspace .
\]
The bottom up dynamic programming will require us to store the error tables 
along at most two leaf to root paths. Thus the required space is,
\[ 2 \sum_{t} R \min \{2^{t},B \} = O(RB(1+\log \frac{n}{B})) \enspace .\]
Since we set $\rho=\epsilon\E/(c n^{1/p}\log n)$, we have 
$\mbox{$R = O((n^{1/p}\log n)/\epsilon)$}$.
\medskip

\begin{theorem}
\label{mainthm}
Algorithm~\ref{alg:apx} is a $O(\epsilon^{-1}B^2n^{1/p}\log^3 n)$ space
algorithm that computes a $(1+\epsilon)$ approximation to the best
$B$-term unrestricted representation of a signal in the Haar
system. Under the $\ell_p$ norm, the algorithm runs in time
$O(\epsilon^{-2}n^{1+2/p}B\log^3 n)$.  Under $\ell_\infty$ the running
time becomes $O(\epsilon^{-2}n\log^2 B\log^3 n)$.
\end{theorem}
\medskip

The extra $B$ factor in the space required by the algorithm accounts
for keeping track of the chosen coefficients.
\smallskip

\subsection{An Improved Algorithm and Analysis}
For large $n$ (compared to $B$), we gain in running time if we change the
rounding scheme given by Lemma~\ref{changebase1}.  The granularity at
which we search for the value of a coefficient will be fine if the
coefficient lies toward the top of the tree, and it will be coarse if
the coefficient lies toward the bottom. The idea is that, for small
$\ell_p$ norms, a mistake in a coefficient high in the tree affects
everyone, whereas mistakes at the bottom are more localized.  This
idea utilizes the strong locality property of the Haar basis.  We
start with the lemma analogous to Lemma~\ref{changebase1}.

\begin{lemma}
\label{changebase3}
Let $\{ y^*_i\}$, $i = (t_i,s)$ be the optimal solution using the
basis set $\{\psib_i\}$ for the reconstruction, i.e., $\hat{f} =
\sum_i y^*_i\psib_i$ and $\| f - \hat{f}\|_p = \E$. Here $t_i$ is the
height of node $i$ in the Haar coefficient tree.  Let $\{y^\rho_i\}$
be the set where each $y^*_i$ is first rounded to the nearest multiple
of $\rho_{t_i} = \epsilon\E / (2B 2^{t_i/p})$ then the resulting value
is rounded to the nearest multiple of $\rho_{t_\text{root}} =
\epsilon\E/(2Bn^{1/p})$. If $f^\rho = \sum_i y^\rho_i\psib_i$ then
$\|f - f^\rho\|_p \leq (1+\epsilon)\E$.
\end{lemma}
\begin{proof}
As in Lemma~\ref{changebase1}, we need to estimate
$\norm{\sum\nolimits_i (y_i^\rho-y_i^*)\psib_i}_p$ but using the new
rounding scheme.  Let $\mathcal{S}$ be the set of indices $i$ such
that $y_i \ne 0$.
\begin{IEEEeqnarray*}{rCl}
\norm{\sum\nolimits_{i\in\mathcal{S}} (y_i^\rho-y_i^*)\psib_i}_p 
& \ \le\  & \sum\nolimits_{i\in\mathcal{S}}\norm{(y_i^\rho-y_i^*)\psib_i}_p \\
&\ \le \ & \sum\nolimits_{i\in\mathcal{S}}(\rho_{t_i} + \rho_{t_\text{root}})\norm{\psib_i}_p \\
&\ \le \ & 2\sum\nolimits_{i\in\mathcal{S}}\rho_{t_i} 2^{t_i/p} \enspace .
\end{IEEEeqnarray*}
The last inequality follows from the fact that $2^{t_i}$ components of
$\psib_i$ are equal to one and the rest are zero. The approximation
hence follows from $|\mathcal{S}| \le B$ and our choices of
$\rho_{t_i}$.
\end{proof}

The granularity of the dynamic programming tables $E[i,\cdot,\cdot]$
is set according to the smallest $\rho_{t_i}$ which is
$\rho_{t_\text{root}} = \epsilon\E/(2Bn^{1/p})$. This allows their
values to align correctly.  More specifically, when a coefficient is
not chosen we compute (see Section~\ref{sec:HaarAlgo})
\[ E[i,v,b] = \min_{b'} E[i_L, v, b'] + E[i_R, v, b-b']\enspace . \]
A value $v$ will that is not outside the range of $E[i_L,\cdot,\cdot]$
and $E[i_R,\cdot,\cdot]$ will be a correct index into these two
arrays.  We gain from this rounding scheme, however, when we are
searching for a value to assign to node $i$.  If $i$ is chosen, we can
search for its value in the range 
$\langle f, \psia_i\rangle \pm 2C_0\E/\rho$ in multiples of $\rho_{t_i}$.  
Hence, as mentioned earlier, the granularity of our search will be
fine for nodes at top levels and coarse for nodes at lower levels.
More formally, if $i$ is chosen, we compute
\[ E[i,v,b] = \min_{r,b'} E[i_L, v+r, b'] + E[i_R, v-r, b-b'-1]\enspace ,\]
where we search for the best $r$ in multiples of $\rho_{t_i}$.
The value $v+r$ (resp.~$v-r$) may not index correctly into
$E[i_L,\cdot,\cdot]$ (resp.~$E[i_R,\cdot,\cdot]$) since 
$\rho_{t_{i}} = 2^{d/p}\rho_{t_{\text{root}}}$ where 
$d = t_{root} - t_i$. Hence, we need to round each value of $r$ we
wish to check to the nearest multiple of
$\rho_{t_{\text{root}}}$. This extra rounding is accounted for in
Lemma~\ref{changebase3}.

Letting $R$ be the number of values each table holds and 
$R_{t_i} = 2C_0\E/\rho_{t_i} + 2$ be the number of entries we search
at node $i$, and using an analysis similar to that of
Section~\ref{sec:algspacetime}, the running time (ignoring constant
factors) becomes,
\begin{align*}
O(\sum_{i=1}^n RR_{t_i}\min\{2^{2t}, B^2\})
&\ =\ O(R \sum_{t=1}^{\log n} \frac{n}{2^t}\frac{B2^{t/p}}{\epsilon} \min\{2^{2t}, B^2\}) \\
&\ =\ O(\frac{nRB}{\epsilon}\left(\sum_{t=1}^{\log B} 2^{t/p+t} + B^2\sum_{t=\log B +1}^{\log n}2^{t/p-t}\right)) \\
&\ =\ O(\frac{nRB}{\epsilon}B^{1+1/p})
\end{align*}
Hence, since $R = O(n^{1/p}B/\epsilon)$ based on the granularity
$\rho_{t_\text{root}}$, the running time for each instance of the
algorithm is $O((nB)^{1+1/p}B^2/\epsilon^2)$.  The space requirement
is the same as that of the simpler algorithm; namely, $O(RB\log n)$.
\smallskip

\begin{theorem}\label{mainthm2}
The above algorithm (with the new rounding scheme) is a
$O(\epsilon^{-1}B^3n^{1/p}\log^2 n)$ space algorithm that computes a
$(1+\epsilon)$ approximation to the best $B$-term unrestricted
representation of a signal in the Haar system under the $\ell_p$ norm.
The algorithm runs in time $O(\epsilon^{-2}(nB)^{1+1/p}B^2\log n)$.
\end{theorem}
\medskip

Again, and as in Theorem~\ref{mainthm}, the extra $B$ factor in the
space requirement accounts for keeping track of the chosen
coefficients, and the extra $\log n$ factor in both the space and time
requirements accounts for the guessing of the error. 

We choose the better of the two algorithms (or rounding schemes) whose
approximation and time and space requirements are guaranteed by
Theorems~\ref{mainthm} and~\ref{mainthm2}.

\section{Extensions}\label{sec:extensions}

\subsection{PTAS for multi-dimensional Haar Systems}

\hide{ %Hiding due to reviewer suggestion.  Also hiding paragraph referencing the figure.
%\begin{figure*}[htb]
\begin{figure}
\begin{center}
\begin{tabular}{lr}
\includegraphics[scale=0.5]{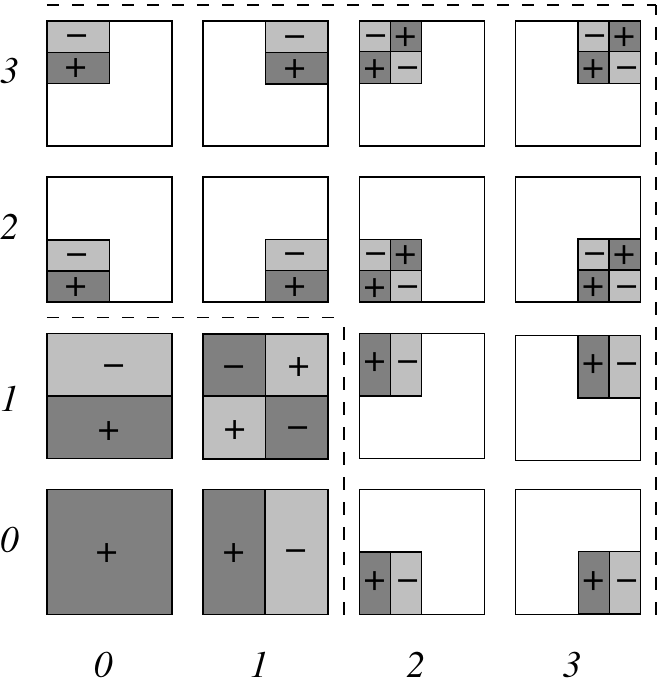} &\hspace{0.5in} 
 \includegraphics[height=1.3in]{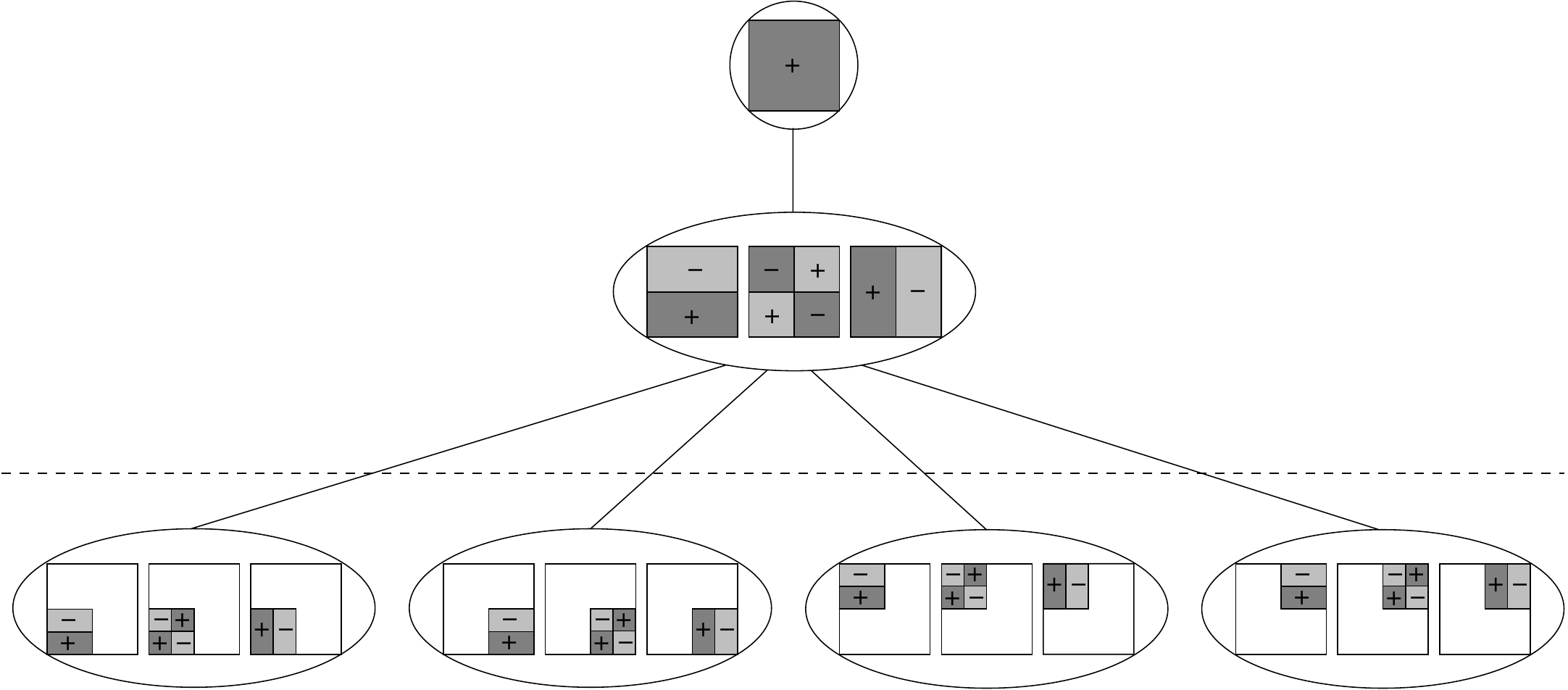} 
\end{tabular}
\end{center}
\caption{Support regions and coefficient tree structure of the
$2$-dimensional Haar basis vectors for a $4\times 4$ data array.
Each node other than the root contains the $3$ wavelet coefficients
having the same support region.
\label{fig:2DHaar}}
\end{figure}
}

Our algorithm and analysis from Section~\ref{apxschemes} extend to
multi-dimensional Haar wavelets when the dimension $D$ is a given
constant.  For $D \ge 2$ define $2^D -1$ mother wavelets (see also
\cite{GG-TODS,GK04})..  For all integers $0\le d < 2^D$ let
\[ \psi^d(x) = \theta^{d_1}(x_1)\theta^{d_2}(x_2)\cdots \theta^{d_D}(x_D) \enspace , \]
where $d_1d_2\ldots d_D$ is the binary representation of $d$ and
$\theta^0 = \phi$, $\theta^1 = \psi$.  For $d = 0$ we obtain the
$D$-dimensional scaling function $\psi^0(x) = \phi(x_1)\phi(x_2)\cdots
\phi(x_D)$. At scale $2^j$ and for $s = (s_1, s_2, \ldots, s_D)$
define
\[ \psi^d_{j,s}(x) = 2^{-Dj/2}\psi^d\left(\frac{x_1 - 2^j s_1}{2^j}, 
  \cdots, \frac{x_D - 2^j s_D}{2^j}\right) \enspace .\] The family
$\{\psi^d_{j,s}\}_{1\le d< 2^D, (j, n)\in \mathbb{Z}^{D+1}}$ is an
orthonormal basis of $L^2(\mathbb{R}^D)$ \cite[Thm.~7.25]{wave2}.
Note that in multi-dimensions we define $\psi^{a,d}_{j,s} =
2^{-Dj/2}\psi^d_{j,s}$, $\psi^{b,d}_{j,s} = 2^{Dj/2}\psi^d_{j,s}$ and
$\phi^{a,d}_{j,s} = 2^{-Dj/2}\psi^0_{j,s}$ which is analogous to
Definition~\ref{def:psi-ab}.  Thus $\norm{\psi^{a,d}_{j,s}}_1 =
\norm{\phi^{a,d}_{j,s}}_1 = 1$ since $\norm{\psi^d_{j,s}}_1 =
2^{Dj}2^{-Dj/2} = 2^{Dj/2}$. Also $\norm{\psi^{b,d}_{j,s}}_\infty =
1$.
\hide{As an illustration of a multi-dimensional basis,
Fig.~\ref{fig:2DHaar} \cite[Fig.~1,2]{GK04} shows the support
regions and signs of the $16$ two-dimensional Haar basis vectors
corresponding to a $4\times 4$ input array.  The figure shows that the
support region of say $\psi^3_{1,(0,0)}$, which corresponds to entry
$(2,2)$ in the array, is the $2^{jD} (j=1, D=2)$ elements comprising
the lower left quadrant of the input array.  Fig.~\ref{fig:2DHaar}
also shows the coefficient tree for this case.} %
Each node in the coefficient tree has 
$2^D$ children and corresponds to
$2^D - 1$ coefficients (assuming the input is a hypercube).  The
structure of the coefficient tree will result in a $O(R^{2^D-1})$
increase in running time over the one-dimensional case where $R =
O(\epsilon^{-1}n^{1/p}\log n)$.

As in Section~\ref{sec:HaarAlgo}, we associate an error array
$E[i,b,v]$ with each node $i$ in the tree where $v$ is the result
of the choices of $i$'s ancestors and $b \le B$ is the number of
coefficients used by the subtree rooted at $i$.  The size of each
table is thus $O(\min\{2^{Dj}, B\} R)$ where $j$ is the level of the
tree to which $i$ belongs.
% The number of nodes in the subtree rooted at i (residing in level j) 
% is 2^{(j-1)D} * (2^D - 1) + 2^D - 1
When computing an entry $E[i,b,v]$ in the table, we need to choose
the best non-zero subset $S$ of the $2^D - 1$ coefficients that belong
to the node and the best assignment of values to these $|S|$
coefficients.  These choices contribute a factor $O((2R)^{2^D-1})$ to
the time complexity. We also have to choose the best partition of the
remaining $b-|S|$ coefficients into $2^D$ parts adding another
$O(B^{2^D})$ factor to the running time.  We can avoid the latter
factor by ordering the search among the node's children as in
\cite{GK04,GG-TODS}.  Each node is broken into $2^D-1$
\emph{subnodes}: Suppose node $i$ has children $c_1, \ldots, c_{2^D}$
ordered in some manner.  Then subnode $i_t$, will have $c_t$ as its
left child and subnode $i_{t-1}$ as its right child. Subnode
$i_{2^D-1}$ will have $c_{2^D-1}$ and $c_{2^D}$ as its children.  Now
all subnode $i_t$ needs to do is search for the best partition of $b$
into $2$ parts as usual.  Specifically, fix $S$ and the values given
to the coefficients in $S$.  For each $v$, $b'$ with $0\le b' \le
\min\{2^{Dj}, b-|S|\}$, each subnode starting from $i_{2^D-1}$
computes the best allotment of $b'$ coefficients to its children.
This process takes $O(R(\min\{2^{Dj}, B\})^2)$ time per subnode.  For
$\ell_\infty$ the bounds are better.  All the error arrays for the
subnodes are discarded before considering the next choice of $S$ and
values assigned to its elements.  Hence, assuming the input is of size
$N$, and since there are $N/2^{Dj}$ nodes per level of the coefficient
tree, the total running time is
\[ 
O\left(\sum_{j=1}^{\frac{\log N}{D}}  
          \frac{N}{2^{Dj}} (2R)^{2^D-1} 2^D  R(\min\{2^{Dj}, B\})^2\right) 
 = O(NBR^{2^D}) 
\]
where we dropped the constant factors involving $D$ in the final
expression.  Finally recall from Section~\ref{sec:HaarAlgo} that we
need to make $O(\log N)$ guesses for the error $\E$.

\subsection{QPTAS for General Compact Systems}

We show a simple dynamic programming algorithm that finds a
$(1+\epsilon)$-approximation to the wavelet synopsis construction
problem under the $\ell_\infty$ norm.  The algorithm uses $g(q,n) =
n^{O(q(\log q + \log\log n))}$ time and space.  Under the $\ell_p$
norm, the algorithm uses $n^{O(q(\log q + \frac{\log n}{p}))}$ time
and space.  We will describe the algorithm for the Daubechies wavelet
under the $\ell_\infty$ norm.  Recall that the Daubechies filters have
$2q$ non-zero coefficients.

For a given subproblem, call an edge an \emph{interface edge} if
exactly one of its endpoints is in the subproblem.  Each interface
edge has a value associated with it which is eventually determined at
a later stage.  We will maintain that each subproblem has at most
$4q\log n$ interface edges. A subproblem has a table $E$ associated
with it where for each $b\le B$ and each configuration $I$ of values
on interface edges, $E[b, I]$ stores the minimum contribution to
the overall error when the subproblem uses $b$ coefficients and the
interface configuration is $I$.  From Lemma~\ref{lemma:summary},
setting $\rho = \epsilon \E/(c_1 q \log n)$ for some suitably large
constant $c_1$, each interface edge can have one of $V =
O(\frac{q^{3/2}\log n}{\epsilon})$ values under the $\ell_\infty$ norm.
Hence, the size of $E$ is bounded by $B V^{4q\log n} = g(q,n)$.

The algorithm starts with an initialization phase that creates the
first subproblem.  This phase essentially flattens the cone-shape of
the coefficient graph, and the only difference between it and later
steps is that it results in one subproblem as opposed to two.  We
select any $2q$ consecutive leaves in the coefficient graph and their
ancestors.  This is at most $2q\log n$ nodes.  We will guess the
coefficients of the optimal solution associated with this set of
nodes.  Again, from Lemma~\ref{lemma:summary}, each coefficient can
take one of $W = O(\frac{q^{3/2}\log n}{\epsilon})$ values under the
$\ell_\infty$ norm.  For each of the $(2W)^{2q\log n} = g(q,n)$
guesses, we will run the second phase of the algorithm.

\begin{figure}
\centering
\includegraphics[scale=0.65]{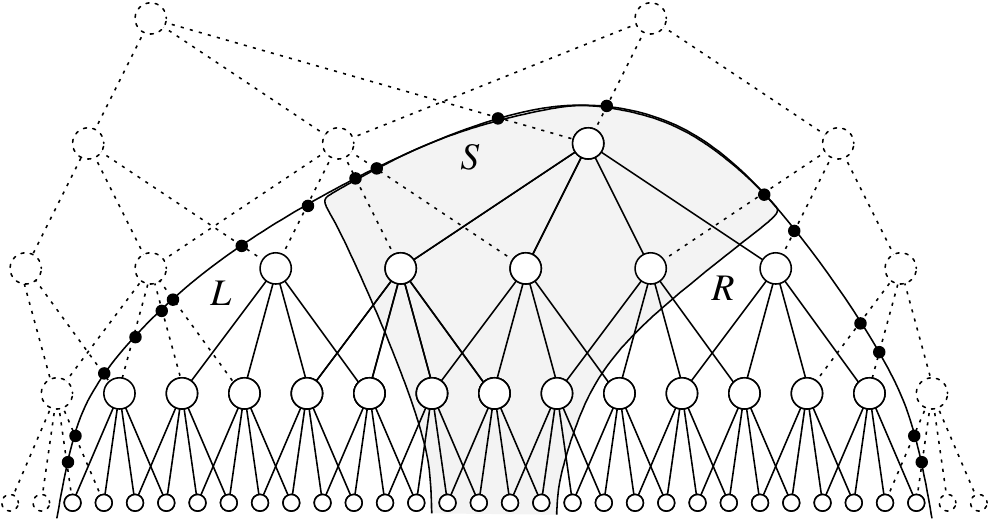}
\caption{An example subproblem.  The shaded nodes belong to the strip
$S$.  The edges crossing the `frontier' are interface edges.}
\label{figure:daub4-qptas}
\end{figure}

In the second phase, given a subproblem $A$, we first select the $2q$
`middle' leaves and their ancestors.  Call this strip of nodes $S$.
Note that $|S| \le 2q\log n$.  The nodes in $S$ break $A$ into two
smaller subproblems $L$ and $R$ (see Fig.~\ref{figure:daub4-qptas}).
Suppose we have $E_L$ and $E_R$, the two error arrays associated
with $L$ and $R$ respectively.  We compute each entry $E_A[b, I]$
as follows. First, we guess the $b'$ non-zero coefficients of the
optimal solution associated with the nodes in $S$ and their values.
Combined with the configuration $I$, these values define a
configuration $I_L$ (resp. $I_R$) for the interface edges of $L$
(resp. $R$) in the obvious way.  Furthermore, they result in an error
$e$ associated with the leaf nodes in $S$.  Hence,
\[ 
E[b,I]  = e\ +\min_{b''}\max\{E_L[b'',I_L],E_R[b-b'-b'',I_R]\}\ .
\]
Therefore, computing each entry in $E$ takes at most
$B(2W)^{2q\log n} = g(q,n)$ time.  The running time of
the algorithm follows.

\begin{theorem}
\label{mainthmG}
We can compute a $(1+\epsilon)$ approximation to the best $B$-term
unrestricted representation of a compact system under the
$\ell_\infty$ norm in time $n^{O(q(\log q + \log \log n))}$.
\end{theorem}

The result also extends to $\ell_p$ norms, but remains a quasi-polynomial
time algorithm. The main point of the above theorem is that the
representation problem is not {\sc Max-SNP-Hard}.

\subsection{Workloads}\label{sec:workloads}
The algorithm and analysis from Section~\ref{apxschemes} also extend
to weighted cases/workloads under the same assumptions as in
\cite{yossi1}. Namely, given $f$ and $\{w_i\}$ where
$\sum_{i=1}^n w_i = 1$ and $0 < w_i \le 1$, we wish to find a solution
$\{z_i\}$ with at most $B$ non-zero coefficients that minimizes,
\[ \left\| f - \sum\nolimits_i z_i\psi_i\right\|_{p,\vect{w}}= 
\left(\sum\nolimits_j w_i^p\left|f(j) - \sum\nolimits_i z_i\psi_i(j)\right|^p\right)^{1/p}\enspace .\] 
Letting $w = \min_i w_i$ and $W = \max_i w_i $, we will show how our
approximation algorithm extends to this case with a factor
$\frac{W}{w}$ increase in its space requirement and a factor
$\left(\frac{W}{w}\right)^2$ increase in running time.

The following three lemmas are analogs of lemmas \ref{changebase1},
\ref{philemma} and \ref{psilemma} respectively.  The first two are
straightforward, but note the factor $W$ in the additive
approximation.
\begin{lemma}
Let $\{ y^*_i\}$ be the optimal solution using the basis set
$\{\psib_i\}$ for the reconstruction, i.e., $\hat{f} = \sum_i
y^*_i\psib_i$ and $\| f - \hat{f}\|_{p,\vect{w}} = \E$. Let
$\{y^\rho_i\}$ be the set where each $y^*_i$ is rounded to the nearest
multiple of $\rho$. If $f^\rho = \sum_i y^\rho_i\psib_i$ then 
$\|f -f^\rho\|_{p,\vect{w}} \leq \E + O(qWn^{1/p}\rho\log n)$.
\end{lemma}

\begin{lemma}
$-C_1\sqrt{q}\frac{\E}{w} \leq \langle f, \phia_{j,s}\rangle - \langle
\hat{f}, \phia_{j,s}\rangle \leq C_1\sqrt{q}\frac{\E}{w}$ for some
constant $C_1$.
\end{lemma}

\begin{lemma}
$-C_0\sqrt{q}\frac{\E}{w}\leq \langle f, \psia_{j,s}\rangle - y^*_i
\leq C_0\sqrt{q}\frac{\E}{w}$ for some constant $C_0$.
\end{lemma}
\begin{proof}
For all $j$ we have $w_j\abs{f(j) - \sum_i y^*_i\psib_i(j)}\leq \E$.
Multiplying by $\abs{\psia_k(j)}$ and summing over all $j$ we get
\begin{align*}
& \sum_j w_j\abs{f(j)\psia_k(j) - \sum_i y^*_i\psib_i(j)\psia_k(j)}\ 
\leq\ \norm{\psia_k}_1\E \\
\Rightarrow\quad &  
w\abs{\sum_j f(j)\psia_k(j) - \sum_i y^*_i\sum_j\psi_i(j)\psi_k(j)}\ 
\leq\ \norm{\psia_k}_1\E \\
\Rightarrow\quad & 
w\abs{\langle f, \psia_k \rangle - y^*_k} \leq \sqrt{2q}\E \enspace ,
\end{align*}
completing the proof.
\end{proof}
Hence, setting $\rho = \epsilon\E/(c q W n^{1/p}\log n)$ for some
suitably large constant $c$, we get the desired approximation with $R$
from the analysis above equal to $O(\E/(w\rho)) =
O(\frac{W}{w}q\epsilon^{-1}n^{1/p}\log n)$.

\subsection{Quality Versus Time}\label{jitter} 
A natural question arises, if we were interested in the restricted
synopses only, can we develop streaming algorithms for them? The answer
reveals a rich tradeoff between synopsis quality and running time.

Observe that if at each node we only consider
either storing the coefficient or $0$, then we can limit the search
significantly. Instead of searching over all $v+r$ to the left and
$v-r$ to the right in the dynamic program (which we repeat below)
\[ \min \left \{ \begin{array}{l}
\min_{r,b'} E[i_L,v+r,b'] + E[i_R,v-r,b-b'-1] \\
\min_{b'} E[i_L,v,b'] + E[i_R,v,b-b']
\end{array} \right. \enspace ,
\]
we only need to search for $r=\langle f, \psia_i\rangle$---observe
that a streaming algorithm can compute $\langle f, \psia_i\rangle$
(See \cite{GKMS01}). However we have to ``round'' $\langle f,
\psia_i\rangle$ to a multiple of $\rho$ since we are storing the table
corresponding to the multiples of $\rho$ between $\langle f,
\phia_i\rangle - C_1\sqrt{q}\E$ and $\langle f, \phia_i\rangle +
C_1\sqrt{q}\E$. We consider {\em the better of rounding up or rounding
down $\langle f, \psia_i\rangle$ to the nearest multiple of
$\rho$}. The running time improves by a factor of $R$ in this case
since in order to compute each entry we are now considering only two
values of $R_i$ (round up/down) instead of the entire set. The overall
running time is $O(RnB)$ in the general case and $O(Rn \log^2 B)$
for the $\ell_\infty$ variants.  The space bound and the approximation
guarantees remain unchanged.  However the guarantee is now against the
synopsis which is restricted to storing wavelet coefficients.

The above discussion sets the ground for investigating a variety of
{\em Hybrid algorithms} where we choose different search strategies
for each coefficient.  We introduced this idea in \cite{GH05} but in
the context of a weaker approximation strategy.  One strategy we
explore in Section~\ref{sec:expt} is to allow the root node to range
over the set $R_1$ while considering the better of rounding up or
rounding down $\langle f, \psia_i\rangle$ to the nearest multiple of
$\rho$ for all other coefficients ($i>1$).  We show that this simple
modification improves on the quality of the restricted synopsis and on
the running time of the unrestricted algorithm.

\section{Best Basis Selection from a Dictionary}\label{sec:best}

In this section we show how our algorithms can be extended to find
representations in certain types of tree-structured dictionaries.
Specifically, the dictionaries we consider are full binary
tree-structured dictionaries composed of compactly supported
wavelets.  Given $f\in\mathbb{R}^n$, $B\in\mathbb{Z}$ and such a
dictionary $\mathcal{D}$, we now wish to find the best $B$-term
representation of $f$ in a basis from $\mathcal{D}$.  Notice that we
seek both the best basis in $\mathcal{D}$ for representing $f$ using
$B$ terms and the best $B$-term representation of $f$ in this
basis. The error of the representation is its $\ell_p$ distance from
$f$.  We show in Theorem~\ref{thm:bi} how our algorithms from the
previous sections can be used to find provably-approximate answers to
this bi-criteria optimization problem.

We start with the description of our tree-structured dictionaries.
Similar to Coiffman and Wickerhauser \cite{CW92}, our dictionaries
will be composed of $O(n\log n)$ vectors, and will contain
$2^{O(\frac{n}{2})}$ bases: equal to the number of cuts in a complete
binary tree.

%%% We call a basis in a dictionary $\mathcal{D}$ a 
%%% \emph{best $B$-basis for $f$} if, over all bases in $\mathcal{D}$, it results
%%% in the least $\ell_p$ error when representing $f$ using $B$ terms.   

Let $a_{(j,p)} = 2^jp$ and let $g_{(j,p)}[t] = {\bf 1}_{[a_{(j,p)},\
a_{(j,p+1)}-1]}[t]$ be the discrete dyadic window that is $1$ in
$[a_{(j,p)},\ a_{(j,p+1)}-1]$ and zero elsewhere.  Each node in $\cal
D$ is labeled by $(j,p)$, $0\le j\le \log n$, $0\le p\le n2^{-j}-1$,
where $j$ is the height of the node in the tree (the root is at height
$\log n$), and $p$ is the number of nodes to its left that are at the
same height in a complete binary tree.  With each node $(j,p)$ we
associate the subspace $\mathcal{W}_{(j,p)}$ of $\mathbb{R}^n$ that
exactly includes all functions $f\in \mathbb{R}^n$ whose support lies
in $g_{(j,p)}$.  Clearly, $\mathcal{W}_{(\log n, 0)} = \mathbb{R}^n$.

Now suppose $\{e_{k,l}\}_{0\le k < l}$, $l > 0$ is an orthonormal
basis for $\mathbb{R}^l$.  Then,
\[ \mathcal{B}_{(j,p)} = 
\left\{\psi_{k,(j,p)}[t] = g_{(j,p)}[t]e_{k,2^j}[t-a_{(j,p)}] \right\}_{0\le k < 2^j} 
\enspace , \]
is an orthonormal basis for $\mathcal{W}_{(j,p)}$.

\begin{proposition}\label{prop:bunion}
For any internal node $(j,p)$ in the dictionary $\cal D$,
$\mathcal{W}_{(j-1,2p)}$ and $\mathcal{W}_{(j-1,2p+1)}$ are
orthogonal, and
\[ \mathcal{W}_{(j,p)} = 
\mathcal{W}_{(j-1,2p)} \oplus \mathcal{W}_{(j-1,2p+1)} 
\enspace .\]
\end{proposition}
\hide{ % Removing proof per reviewer's suggestion
\begin{proof}
The two spaces $\mathcal{W}_{(j-1,2p)}$ and $\mathcal{W}_{(j-1,2p+1)}$
are orthogonal since $\forall k, \langle \psi_{k,(j-1,2p)},
\psi_{k,(j-1,2p+1)}\rangle = 0$ (as their supports do not overlap).
Now any $f\in\mathcal{W}_{(j,p)}$ can be written as a combination of
two blocks: $f[t] = g_{(j-1,2p)}[t]f[t] + g_{(j-1,2p+1)}[t]f[t]$; and
the first block (resp.~second block) can be represented using
$\{\psi_{k,(j-1,2p)}\}_k$ (resp.~$\{\psi_{k,(j-1,2p+1)}\}_k$).
\end{proof}
\medskip
}%
We can thus construct an orthonormal basis of $\mathcal{W}_{(j,p)}$
via a union of orthonormal bases of $\mathcal{W}_{(j-1,2p)}$ and
$\mathcal{W}_{(j-1,2p+1)}$. 

\begin{corollary}
Let $\{(j_i, p_i)\}$ be the set of nodes corresponding to a cut in the
dictionary tree.  We have,
\[ \bigoplus\nolimits_i \mathcal{W}_{(j_i,p_i)} = \mathbb{R}^n \enspace . \] 
\end{corollary}
Hence, there are $O(2^{n/2})$ bases in our dictionary.
\medskip

The main result of this section follows.  We prove it under the
$\ell_\infty$ error measure.  The argument is extended to general
$\ell_p$ error measures in a straightforward manner.

\begin{theorem}\label{thm:bi}
If $A$ is an (streaming) algorithm that achieves a $C$-approximation for the
$B$-term representation problem under $\ell_\infty$ (for any wavelet
included in the dictionary $\mathcal{D}$), then $A$ is a (streaming)
$C$-approximation for the bi-criteria representation problem.
\end{theorem}
\begin{proof}
Let $E_{(j,p)}[b]$ be the minimum contribution to the overall error
(as computed by $A$) from representing the block $g_{(j,p)}[t]f[t]$
using $b$ vectors from a basis of $\mathcal{W}_{(j,p)}$.  Call the
basis that achieves this error the \emph{best basis for
$\mathcal{W}_{(j,p)}$} and denote it by $\mathcal{O}_{(j,p)}[b]$.  By
Proposition \ref{prop:bunion} there are $O(2^{2^{j-1}})$ possible
bases for the space $\mathcal{W}_{(j,p)}$ in $\mathcal{D}$. Now if
$(j,p)$ is a leaf node, then $E_{(j,p)}[b] = A(g_{(j,p)} \odot f,
\mathcal{B}_{(j,p)}, b)$, which is the error resulting from
representing the block $g_{(j,p)}[t]f[t]$ using $b$ vectors from the
basis $\mathcal{B}_{(j,p)}$.  Otherwise, if $(j,p)$ is an internal
node, $E_{(j,p)}[b]$ equals
\[  \min\left\{ 
\begin{aligned}
& \min_{0\le b' \le b} \max\left\{ E_{(j-1,2p)}[b'],\  E_{(j-1,2p+1)}[b-b']  \right\} \\
& A(g_{(j,p)} \odot f, \mathcal{B}_{(j,p)}, b) 
\end{aligned}
\right. \enspace ,
\]
and 
\[
\mathcal{O}_{(j,p)}[b] =  \left\{
\begin{aligned}
&\mathcal{B}_{(j,p)} 
\quad\text{if } E_{(j,p)}[b] = A(g_{(j,p)} \odot f, \mathcal{B}_{(j,p)}, b)\\
&\mathcal{O}_{(j-1,2p)}[b_{(j,p)}] \cup \mathcal{O}_{(j-1,2p-1)}[b-b_{(j,p)}] 
\quad\text{else,} 
\end{aligned}
\right. 
\]
where $b_{(j,p)}$ is the argument that minimizes the top expression in
$ E_{(j,p)}[b]$.
% $b_{(j,p)} = \argmin_{0\le b' \le b} \max\left\{ E_{(j-1,2p)}[b'],\  E_{(j-1,2p+1)}[b-b']  \right\}$.
\medskip

Suppose $\opt$ chooses the cut $\{(j_o, p_o)\}$ with the corresponding
partition $\{b_o\}$ of $B$ and we choose the cut $\{(j_i, p_i)\}$ with
partition $\{b_i\}$.  By the dynamic program above we have,
\begin{subequations}
\begin{align}
&\max_i A(g_{(j_i,p_i)} \odot f, \mathcal{B}_{(j_i,p_i)}, b_i)  = \max_i  E_{(j_i,p_i)}[b_i] \\
&\quad\quad \ \le \max_o E_{(j_o,p_o)}[b_o] \label{approx:1}\\
&\quad\quad \ \le \max_o A(g_{(j_o,p_o)} \odot f, \mathcal{B}_{(j_o,p_o)}, b_o) \label{approx:2}\\
&\quad\quad \ \le C\max_o\opt(g_{(j_o,p_o)} \odot f, \mathcal{B}_{(j_o,p_o)}, b_o) \label{approx:3} \\
&\quad\quad \ = C\ \opt  \label{approx:4} \enspace , 
\end{align}
\end{subequations}
where \eqref{approx:1} follows from the fact that our dynamic program
chooses the best cut and corresponding partition of $B$ among all
possible cuts and partitions based on the errors computed by algorithm
$A$; \eqref{approx:2} follows from the definition of our dynamic
programming table entries $E_{(j,p)}[b]$; \eqref{approx:2} follows
from the assumption that $A$ is a $C$-approximation algorithm; and
\eqref{approx:4} follows from the optimal substructure property of our
problem.
\end{proof}

\section{Comparing Restricted and Unrestricted optimizations}
\label{sec:expt}
We consider two issues in
this section, namely 
(i) the quality of the unrestricted version vis-a-vis
the restricted optimum solution and (ii) the running times of the
algorithms.  We will restrict our experiments to the $\ell_\infty$ norm.

\subsection{The algorithms}
All experiments reported in this section were performed on
a 2 CPU Pentium-III 1.4 GHz with 2GB of main memory, 
running Linux. All algorithms were implemented using 
version 3.3.4 of the gcc compiler. 

We show the performance figures of the following schemes:
\begin{itemize}
\item[] {\bf REST} This characterizes the algorithms for the {\em
restricted} version of the problem. This is the $O(n^2)$ time $O(n)$
space algorithm in \cite{G05} (see also \cite{GK04,muthu-wave,MU}).

\item[] {\bf UNREST} This is the streaming algorithm for the {\em full
    general version} described in Algorithm~\ref{alg:apx} based on the
    discussion in Section~\ref{apxschemes}\footnote{The
    implementation is available from \url{http://www.cis.upenn.edu/~boulos}.}.

\item[] {\bf HYBRID} This is the streaming hybrid algorithm proposed in
Section~\ref{jitter}.
\end{itemize}

Note that the UNREST and HYBRID algorithms are not the additive
approximation algorithms in \cite{GH05} (although we kept the same
names).

\subsection{The Data Sets}
\label{sec:expsyndata}
We chose a synthetic data set to showcase the point made in the
introduction about the sub-optimality of the restricted
versions. Otherwise we use a publicly available real life data set for
our experiment.
 
\begin{itemize} 
\item {\bf Saw:} This is a periodic data set with a line repeated 8
  times, with $2048$ values total. 
  \hide{The data set is shown in Fig.~\ref{fig:dataset1}. }%
\item {\bf DJIA data set:} We used the Dow-Jones Industrial
Average (DJIA) data set available at StatLib\footnote{ See
http://lib.stat.cmu.edu/datasets/djdc0093.} that contains Dow-Jones
Industrial Average (DJIA) closing values from 1900 to 1993. There were
a few negative values (e.g. $-9$), which we removed.  We focused on
prefixes of the data set of sizes up to $16384$.  
\hide{The data set is shown in Fig.~\ref{fig:dataset2}.}%
\end{itemize}
\hide{
\begin{figure}
\centering
\includegraphics[width=3in]{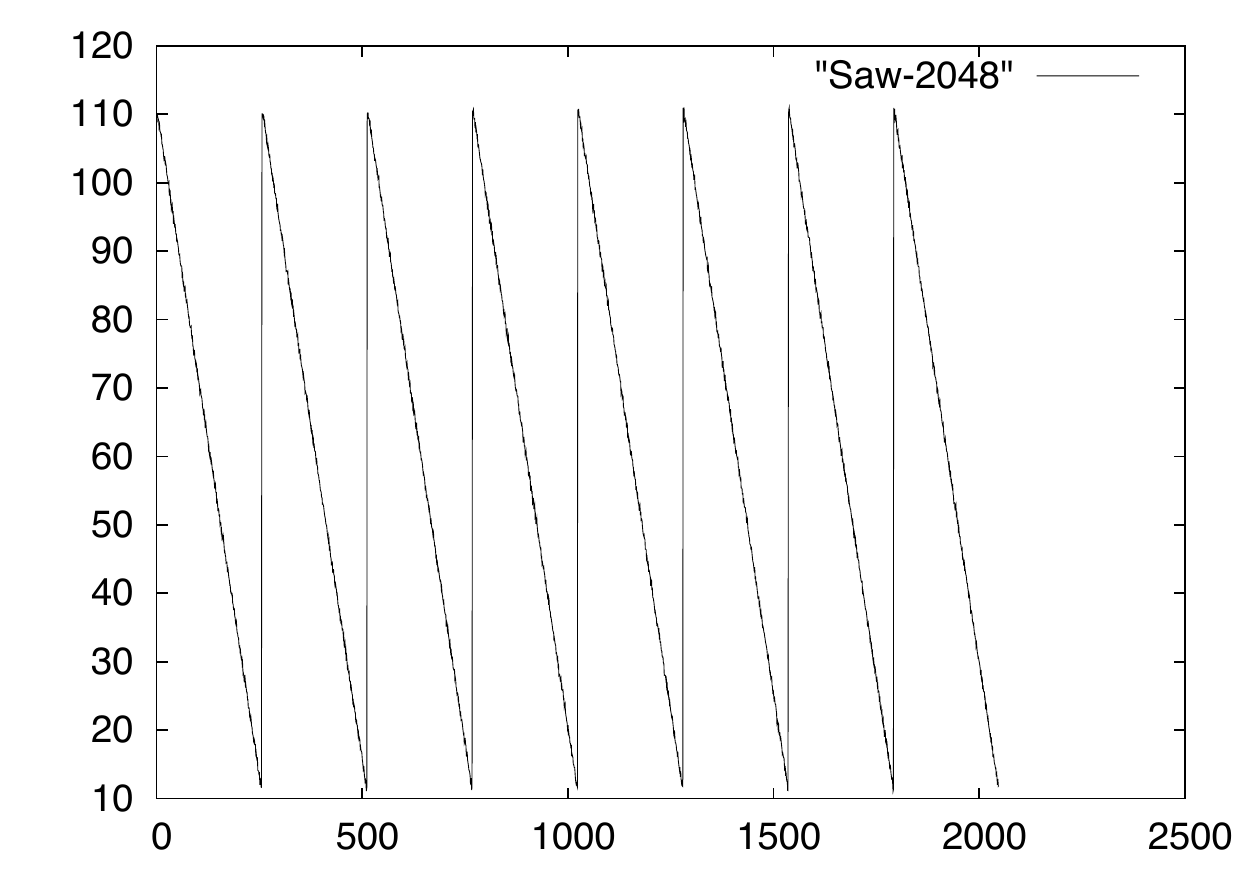} 
\caption{The Saw data set}
\label{fig:dataset1}
\end{figure}
\begin{figure}
\centering
\includegraphics[width=3in]{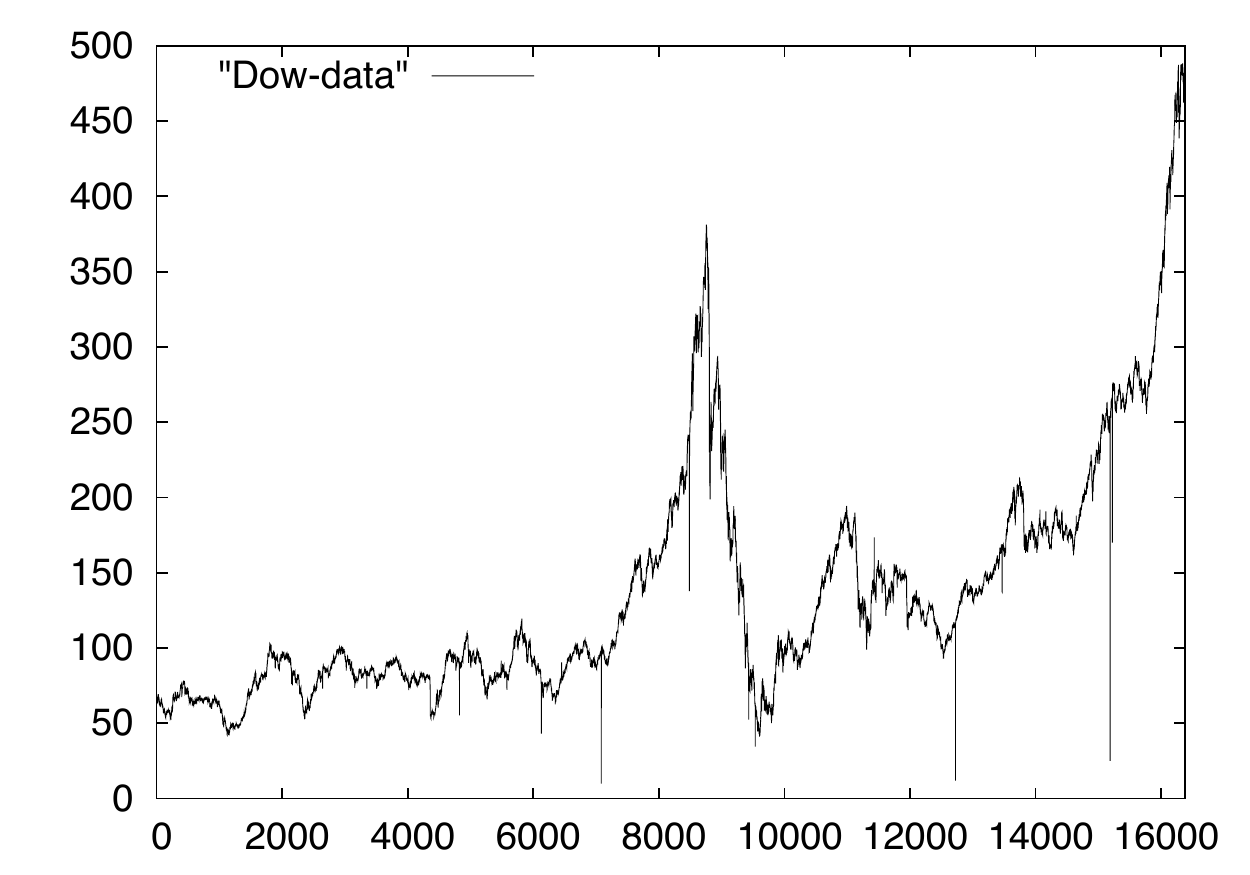} 
\caption{The DJIA data set}
\label{fig:dataset2}
\end{figure}
}%

\subsection{Quality of Synopsis}
The $\ell_\infty$ errors as a function of $B$ are shown in
figures~\ref{fig:errors1} and~\ref{fig:errors2}. The $\epsilon$ in the
approximation algorithms UNREST and HYBRID was set to $1$. All the
algorithms gave very similar synopses for the Saw data and had almost
the same errors. In case of the Dow data we show the range $B=5$
onward since the maximum value is $\sim 500$ and the large errors for
$B<5$ (for all algorithms) bias the scale making the differences in
the more interesting ranges not visible. The algorithm REST has more
than $20\%$ worse error compared to UNREST or requires over $35\%$
more coefficients to achieve the same error (for most error values).
The HYBRID algorithm performs consistently in the middle.
\begin{figure}
\centering
\subfigure[Error for the Saw data set ($n=2048$)]{\label{fig:errors1}
\begin{minipage}[t]{2.83in}
\centering \includegraphics[width=2.83in]{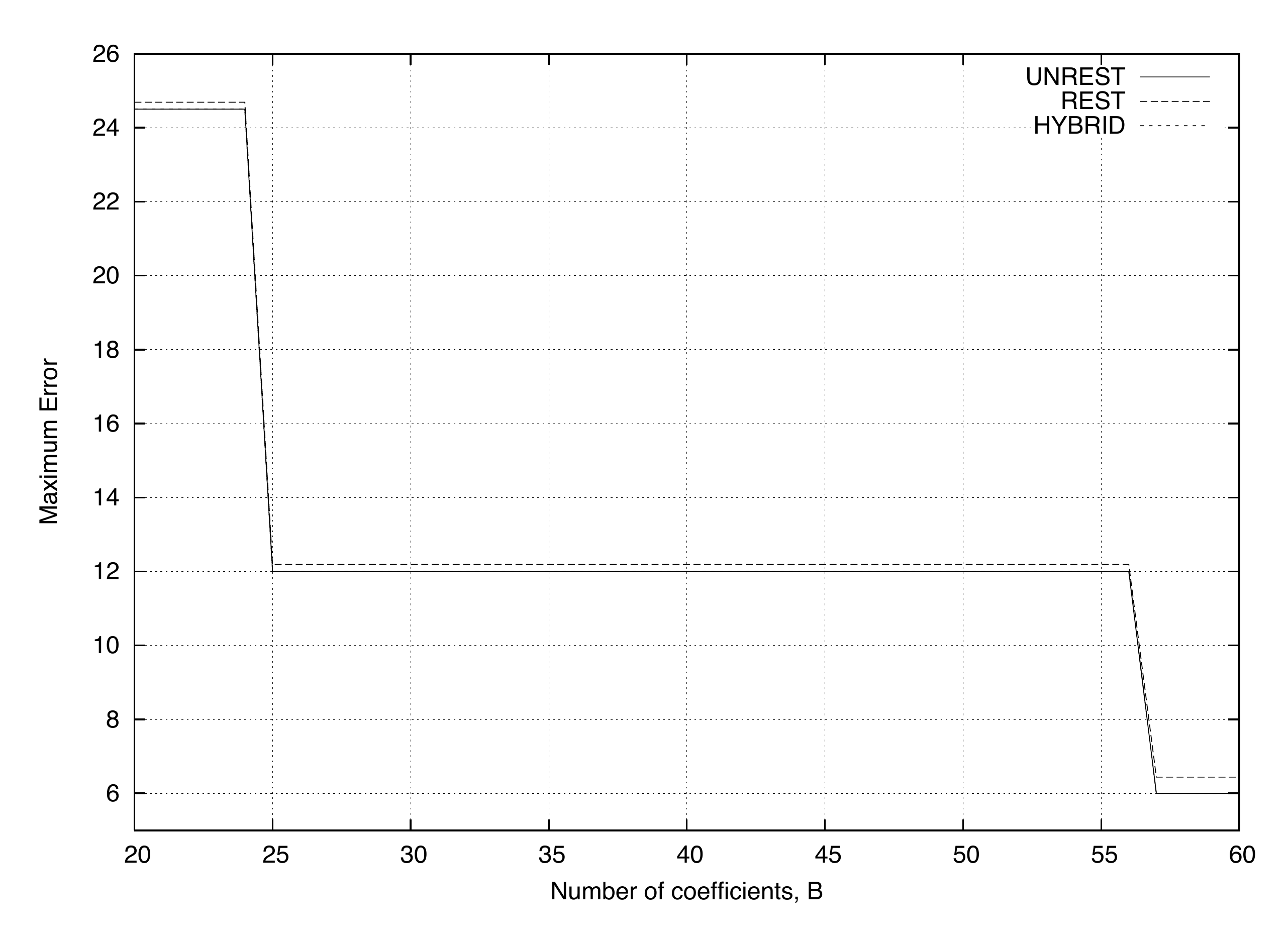}
\end{minipage}
} \subfigure[Error for the Dow data set ($n= 16384 $)]{\label{fig:errors2}
\begin{minipage}[t]{2.83in}
\centering \includegraphics[width=2.83in]{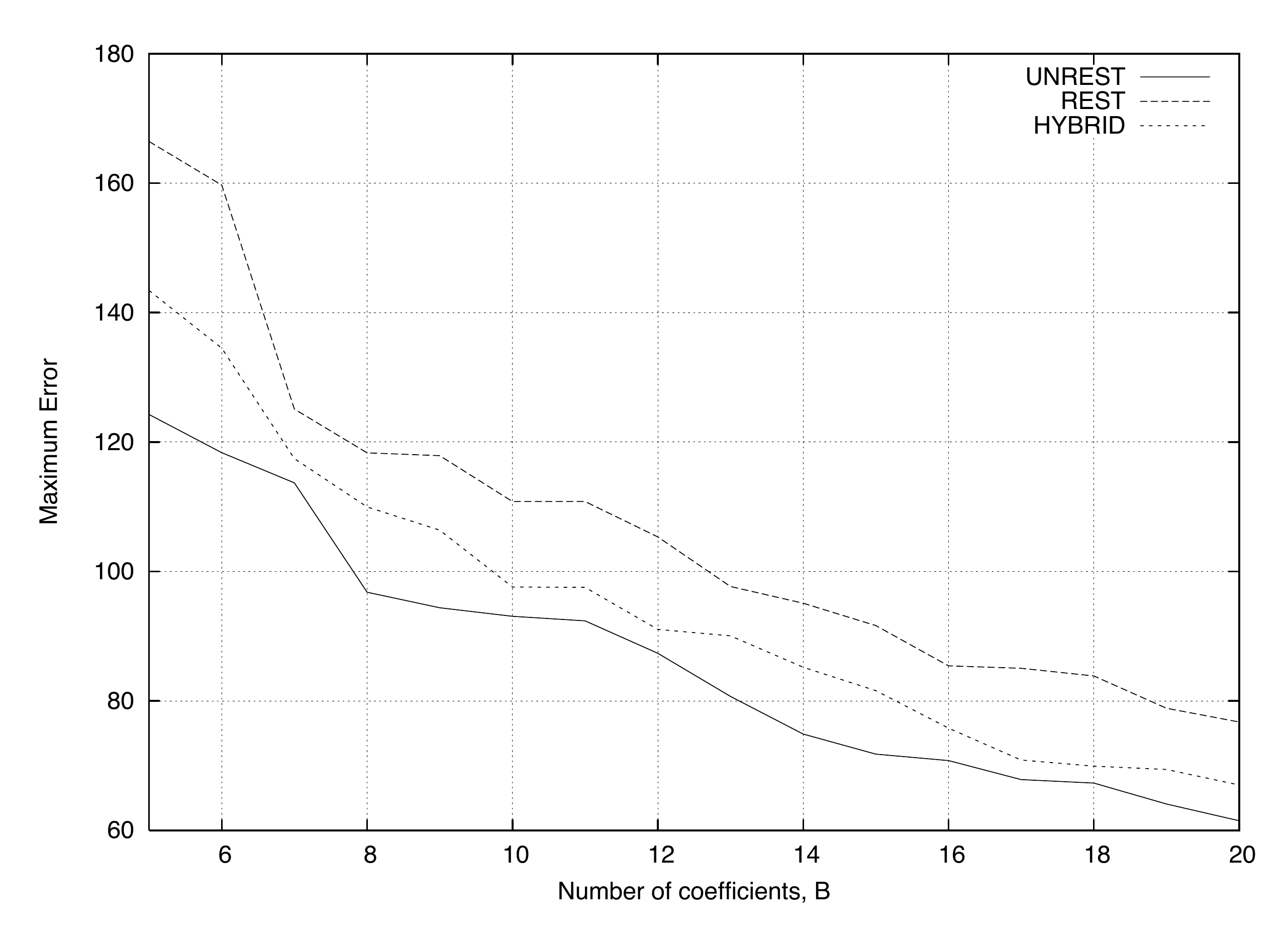}
\end{minipage}} 
\caption{The $\ell_\infty$ error of the three algorithms, UNREST, REST,
and HYBRID for the two data sets.}
\end{figure}
\hide{ % The figures presented seperately
\begin{figure}
\centering
\includegraphics[width=3in]{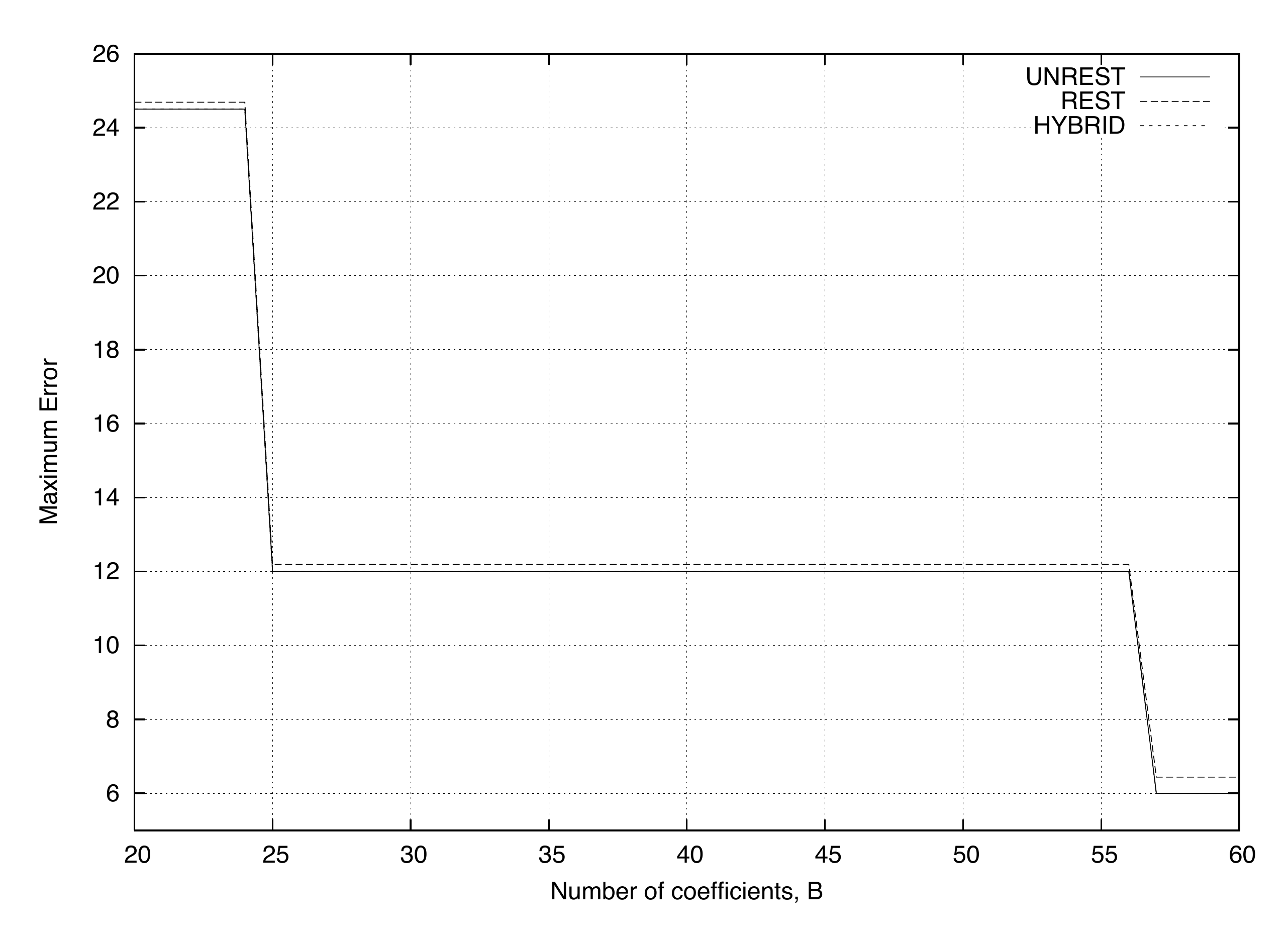} 
\caption{$\ell_\infty$ Error of the three algorithms, UNREST, REST,
and HYBRID for the Saw data set ($n=2048$).}
\label{fig:errors1}
\end{figure}
\begin{figure}
\centering
\includegraphics[width=3in]{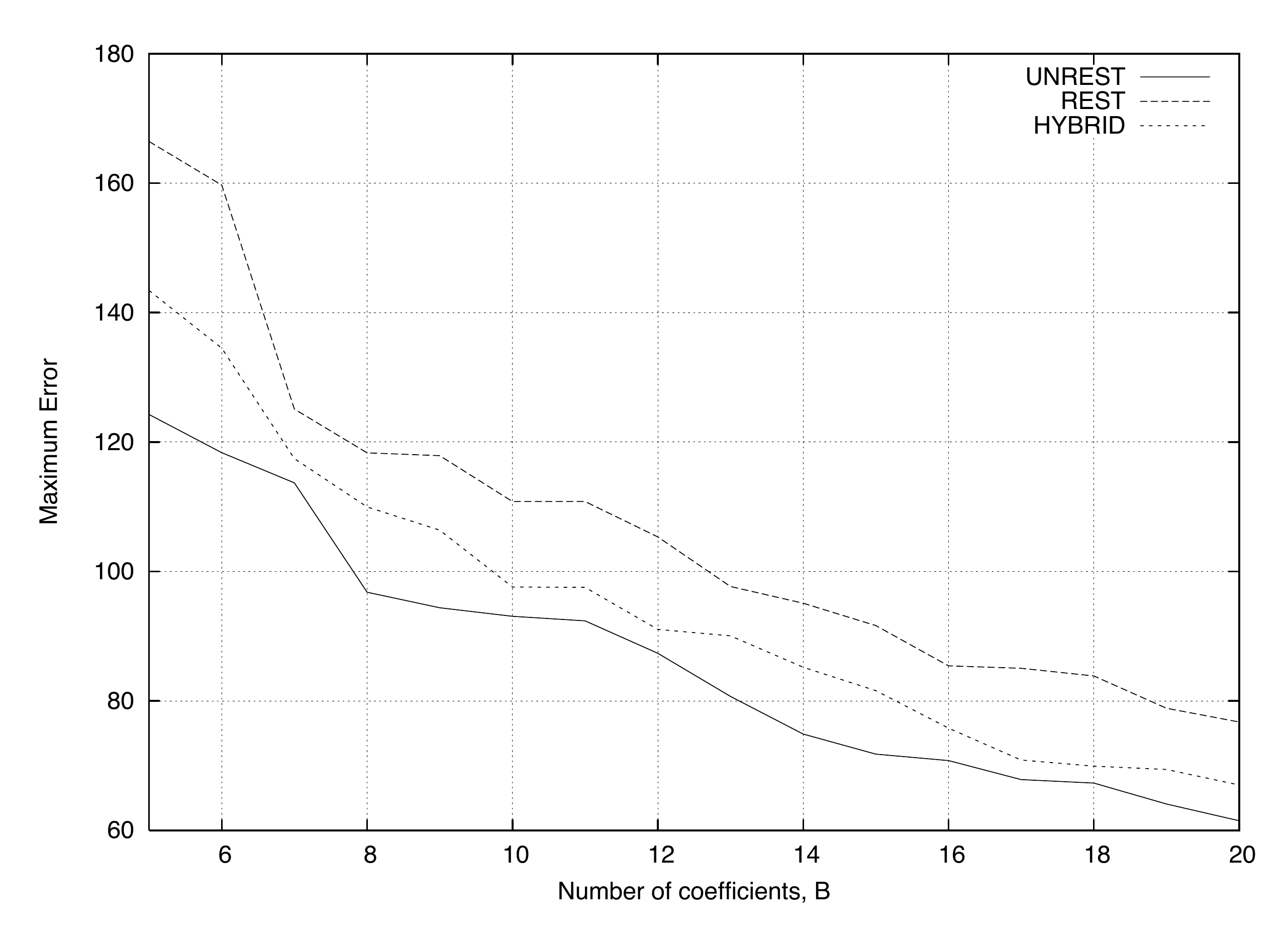} 
\caption{$\ell_\infty$ Error of the three algorithms, UNREST, REST,
and HYBRID for the Dow data set ($n=16384$)}
\label{fig:errors2}
\end{figure}
}

\subsection{Running Times}
\begin{figure}
\centering
\includegraphics[width=3in]{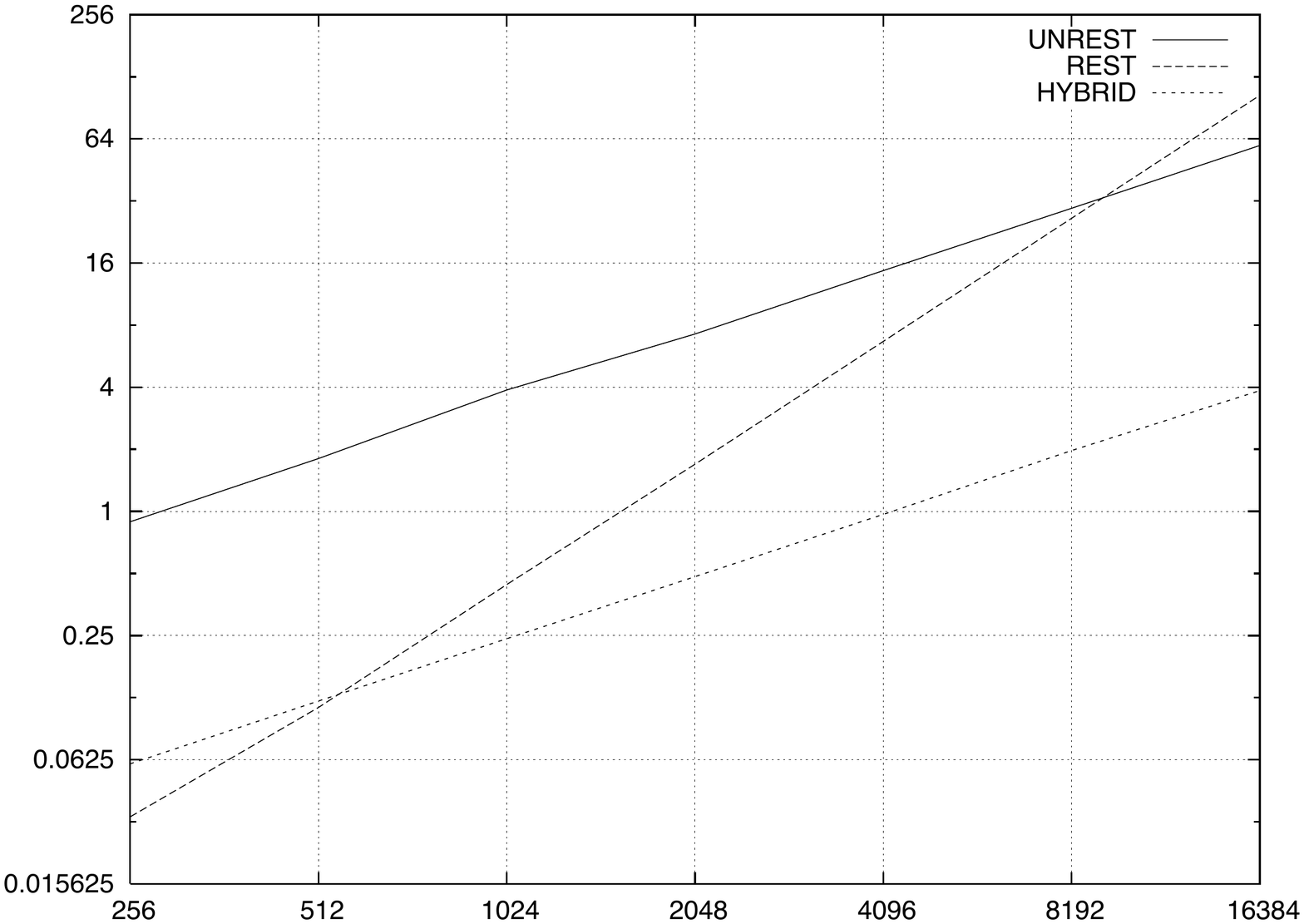} 
\caption{Running times for prefixes of the Dow data set}
\label{fig:times}
\end{figure}

Figure~\ref{fig:times} shows the running times of the algorithms as
the prefix size $n$ is varied for the Dow data. As mentioned above
$\epsilon$ was set to $1$. The grid in the log-log plot helps us
clearly identify the quadratic nature of REST. The algorithms UNREST
and HYBRID behave linearly as is expected from streaming algorithms.
Given its speed and quality, the HYBRID algorithm seems to be the best
choice from a practical perspective.

\bibliographystyle{IEEEtran} 
\bibliography{hist}

\begin{thebibliography}{10}
\providecommand{\url}[1]{#1}
\csname url@rmstyle\endcsname
\providecommand{\newblock}{\relax}
\providecommand{\bibinfo}[2]{#2}
\providecommand\BIBentrySTDinterwordspacing{\spaceskip=0pt\relax}
\providecommand\BIBentryALTinterwordstretchfactor{4}
\providecommand\BIBentryALTinterwordspacing{\spaceskip=\fontdimen2\font plus
\BIBentryALTinterwordstretchfactor\fontdimen3\font minus
  \fontdimen4\font\relax}
\providecommand\BIBforeignlanguage[2]{{%
\expandafter\ifx\csname l@#1\endcsname\relax
\typeout{** WARNING: IEEEtran.bst: No hyphenation pattern has been}%
\typeout{** loaded for the language `#1'. Using the pattern for}%
\typeout{** the default language instead.}%
\else
\language=\csname l@#1\endcsname
\fi
#2}}

\bibitem{GH06}
S.~Guha and B.~Harb, ``Approximation algorithms for wavelet transform coding of
  data streams,'' in \emph{SODA '06: Proceedings of the seventeenth annual
  ACM-SIAM symposium on Discrete algorithm}.\hskip 1em plus 0.5em minus
  0.4em\relax New York, NY, USA: ACM Press, 2006, pp. 698--707.

\bibitem{schmidt}
E.~Schmidt, ``Zur theorie der linearen und nichtlinearen integralgleichungen -
  i,'' \emph{Math. Annalen}, vol.~63, pp. 433--476, 1907.

\bibitem{wave2}
S.~Mallat, \emph{A wavelet tour of signal processing}.\hskip 1em plus 0.5em
  minus 0.4em\relax Academic Press, 1999.

\bibitem{wave3}
I.~Daubechies, \emph{Ten lectures on wavelets}.\hskip 1em plus 0.5em minus
  0.4em\relax Philadelphia, PA, USA: Society for Industrial and Applied
  Mathematics, 1992.

\bibitem{devoreWavelets}
R.~DeVore, B.~Jawerth, and V.~A. Popov, ``Compression of wavelet
  decompositions,'' \emph{Amer. J. Math.}, vol. 114, pp. 737--785, 1992.

\bibitem{wavei}
C.~E. Jacobs, A.~Finkelstein, and D.~H. Salesin, ``Fast multiresolution image
  querying,'' \emph{Computer Graphics}, vol.~29, no. {Annual Conference
  Series}, pp. 277--286, 1995.

\bibitem{cohen97importance}
\BIBentryALTinterwordspacing
A.~Cohen, I.~Daubechies, O.~Guleryuz, and M.~Orchard, ``On the importance of
  combining wavelet-based non-linear approximation in coding strategies,''
  \emph{IEEE Transactions on Information Theory}, vol.~48, no.~7, pp.
  1895--1921, 2002. [Online]. Available:
  \url{citeseer.ist.psu.edu/article/cohen97importance.html}
\BIBentrySTDinterwordspacing

\bibitem{devore}
R.~DeVore, ``Nonlinear approximation,'' \emph{Acta Numerica}, pp. 1--99, 1998.

\bibitem{T03}
V.~N. Temlyakov, ``Nonlinear methods of approximation,'' \emph{Foundations of
  Computational Mathematics}, vol.~3, pp. 33--107, 2003.

\bibitem{MVW98}
Y.~Matias, J.~S. Vitter, and M.~Wang, ``{ Wavelet-Based Histograms for
  Selectivity Estimation},'' \emph{Proc. of ACM SIGMOD}, 1998.

\bibitem{CHV}
O.~Chapelle, P.~Haffner, and V.~Vapnik, ``Support vector machines for
  histogram-based image classification,'' \emph{IEEE transactions on Neural
  Networks}, vol.~10, no.~5, pp. 1055--1064, 1999.

\bibitem{GG-TODS}
M.~N. Garofalakis and P.~B. Gibbons, ``Probabilistic wavelet synopses,''
  \emph{ACM TODS}, vol.~29, pp. 43--90, 2004.

\bibitem{GK05}
M.~Garofalakis and A.~Kumar, ``Wavelet synopses for general error metrics,''
  \emph{ACM Trans. Database Syst.}, vol.~30, no.~4, pp. 888--928, 2005.

\bibitem{muthu-wave}
S.~Muthukrishnan, ``Nonuniform sparse approximation using haar wavelet basis,''
  \emph{DIMACS TR 2004-42}, 2004.

\bibitem{MU}
Y.~Matias and D.~Urieli, ``Personal communication,'' 2004.

\bibitem{yossi1}
------, ``Optimal workload-based weighted wavelet synopses,'' \emph{Proc. of
  ICDT}, pp. 368--382, 2005.

\bibitem{G05}
S.~Guha, ``Space efficiency in synopsis construction problems,'' \emph{Proc. of
  VLDB Conference}, 2005.

\bibitem{GK04}
M.~Garofalakis and A.~Kumar, ``Deterministic wavelet thresholding for maximum
  error metric,'' \emph{Proc. of PODS}, 2004.

\bibitem{T98}
V.~N. Temlyakov, ``The best $m$-term representation and greedy algorithms,''
  \emph{Advances in Computational Mathematics}, vol.~8, pp. 249--265, 1998.

\bibitem{DKT98}
R.~A. DeVore, S.~V. Konyagin, and V.~N. Temlyakov, ``Hyperbolic wavelet
  approximation,'' \emph{Constructive Approximation}, vol.~14, pp. 1--26, 1998.

\bibitem{CW92}
R.~R. Coifman and M.~V. Wickerhauser, ``Entropy-based algorithms for best basis
  selection,'' \emph{IEEE Transactions on Information Theory}, vol.~38, no.~2,
  pp. 713--718, 1992.

\bibitem{MallatNPhrad}
G.~Davis, S.~Mallat, and M.~Avellaneda, ``Adaptive greedy approximation,''
  \emph{Journal of Constructive Approximation}, vol.~13, pp. 57--98, 1997.

\bibitem{redundant}
A.~C. Gilbert, S.~Muthukrishnan, and M.~Strauss, ``Approximation of functions
  over redundant dictionaries using coherence,'' \emph{Proc. of SODA}, pp.
  243--252, 2003.

\bibitem{GKMS01}
A.~C. Gilbert, Y.~Kotidis, S.~Muthukrishnan, and M.~Strauss, ``Optimal and
  approximate computation of summary statistics for range aggregates,'' in
  \emph{Proc. of ACM PODS}, 2001.

\bibitem{GGIKMS02}
A.~C. Gilbert, S.~Guha, P.~Indyk, Y.~Kotidis, S.~Muthukrishnan, and M.~Strauss,
  ``Fast, small-space algorithms for approximate histogram maintenance,'' in
  \emph{Proc. of ACM STOC}, 2002.

\bibitem{F98}
U.~Feige, ``A threshold of $\ln n$ for approximating set cover,'' \emph{J.
  ACM}, vol.~45, no.~4, pp. 634--652, 1998.

\bibitem{GMMO00}
S.~Guha, N.~Mishra, R.~Motwani, and L.~O'Callaghan, ``Clustering data
  streams,'' \emph{Proceedings of the Symposium on Foundations of Computer
  Science (FOCS)}, pp. 359--366, 2000.

\bibitem{GIMS02}
S.~Guha, P.~Indyk, S.~Muthukrishnan, and M.~Strauss, ``Histogramming data
  streams with fast per-item processing,'' in \emph{Proc. of ICALP}, 2002.

\bibitem{keogh01}
E.~Keogh, K.~Chakrabati, S.~Mehrotra, and M.~Pazzani, ``{Locally Adaptive
  Dimensionality Reduction for Indexing Large Time Series Databases},''
  \emph{Proc. of ACM SIGMOD, Santa Barbara}, Mar. 2001.

\bibitem{CKMP02}
K.~Chakrabarti, E.~J. Keogh, S.~Mehrotra, and M.~J. Pazzani, ``Locally adaptive
  dimensionality reduction for indexing large time series databases,''
  \emph{ACM TODS}, vol.~27, no.~2, pp. 188--228, 2002.

\bibitem{GKS01}
S.~Guha, N.~Koudas, and K.~Shim, ``{Data Streams and Histograms},'' in
  \emph{Proc. of STOC}, 2001.

\bibitem{I03}
Y.~E. Ioannidis, ``The history of histograms (abridged),'' \emph{Proc. of VLDB
  Conference}, pp. 19--30, 2003.

\bibitem{Harb07}
B.~Harb, ``Algorithms for linear and nonlinear approximation of large data,''
  Ph.D. dissertation, University of Pennsylvania, 2007.

\bibitem{Haar10}
A.~Haar, ``Zur theorie der orthogonalen funktionen-systeme,'' \emph{Math.
  Ann.}, vol.~69, pp. 331--371, 1910.

\bibitem{Mallat89}
S.~Mallat, ``Multiresolution approximations and wavelet orthonormal bases of
  $l^2(\mathbb{R})$,'' \emph{Trans. Amer. Math. Soc.}, vol. 315, pp. 69--87,
  September 1989.

\bibitem{Meyer92}
Y.~Meyer, \emph{Wavelets and operators}, ser. Advanced mathematics.\hskip 1em
  plus 0.5em minus 0.4em\relax Cambridge University Press, 1992.

\bibitem{D88}
I.~Daubechies, ``Orthonormal bases of compactly supported wavelets,''
  \emph{Comm. Pure. Appl. Math.}, vol.~41, pp. 909--996, 1988.

\bibitem{christopoulos2000jsi}
C.~Christopoulos, A.~Skodras, and T.~Ebrahimi, ``{The JPEG2000 still image
  coding system: an overview},'' \emph{Consumer Electronics, IEEE Transactions
  on}, vol.~46, no.~4, pp. 1103--1127, 2000.

\bibitem{GKS04}
S.~Guha, C.~Kim, and K.~Shim, ``{XWAVE}: Optimal and approximate extended
  wavelets for streaming data,'' \emph{Proceedings of VLDB Conference}, 2004.

\bibitem{uviwave}
\BIBentryALTinterwordspacing
S.~G. Sanchez, N.~G. Prelcic, and S.~J.~G. Galan, ``Uvi\_{W}ave version
  3.0---{W}avelet toolbox for use with {\sc {m}atlab}.'' [Online]. Available:
  \url{citeseer.ist.psu.edu/672431.html}
\BIBentrySTDinterwordspacing

\bibitem{kokinshu}
H.~McCullough, \emph{Kokin Wakashu: The First Imperial Anthology of Japanese
  Poetry}.\hskip 1em plus 0.5em minus 0.4em\relax Palo Alto: Stanford
  University Press, 1984, translated from Japanese.

\bibitem{GVL89}
G.~Golub and C.~V. Loan, \emph{Matrix Computations}.\hskip 1em plus 0.5em minus
  0.4em\relax Johns Hopkins University Press, 1989.

\bibitem{GH05}
S.~Guha and B.~Harb, ``Wavelet synopsis for data streams: minimizing
  non-euclidean error,'' in \emph{KDD '05: Proceeding of the eleventh ACM
  SIGKDD international conference on Knowledge discovery in data mining}.\hskip
  1em plus 0.5em minus 0.4em\relax New York, NY, USA: ACM Press, 2005, pp.
  88--97.

\end{thebibliography}

\end{document}